\documentclass[journal]{IEEEtran}
\IEEEoverridecommandlockouts
\usepackage{cite}
\usepackage{amsmath,amssymb,amsfonts}
\usepackage{booktabs}
\usepackage[table]{xcolor}
\usepackage{makecell}
\usepackage{algorithm}
\usepackage{algorithmic}
\usepackage{graphicx}
\usepackage{textcomp}
\usepackage{hyperref}
\usepackage{xcolor}
\usepackage[dvipsnames]{xcolor}
\usepackage{amssymb}
\usepackage{subcaption}
\usepackage{comment}
\usepackage{enumitem}
\usepackage{url}
\usepackage{textcomp}
\usepackage{bbm}
\usepackage{pifont}
\newcommand{\cmark}{\ding{51}}
\newcommand{\xmark}{\ding{55}}

\def\BibTeX{{\rm B\kern-.05em{\sc i\kern-.025em b}\kern-.08em
    T\kern-.1667em\lower.7ex\hbox{E}\kern-.125emX}}
\begin{document}

\title{Dual-Graph Multi-Agent Reinforcement Learning for Handover Optimization}

\author{M. Salvatori, F. Vannella, S. Macaluso, S. E. Trevlakis, C. Segura Perales, \\ J. Suarez-Varela, A.-A. A. Boulogeorgos, and I. Arapakis 
\thanks{M. Salvatori, F. Vannella, S. Macaluso, C. Segura Perales, J. Suarez-Varela, and I. Arapakis are with Telefónica Research, Barcelona, 08019, Spain (e-mail: \{matteo.salvatori, filippo.vannella, sebastian.macaluso, carlos.seguraperales, jose.suarez-varela, ioannis.arapakis\}@telefonica.com).}
\thanks{S. E. Trevlakis is with Innocube PC, 55534 Thessaloniki, Greece, (e-mail: trevlakis@innocube.org)}
\thanks{A.-A. A. Boulogeorgos is with the Department of Electrical and Computer Engineering, Democritus University of Thrace, Xanthi, 67100, Greece (e-mail: al.boulogeorgos@ieee.org).}
\thanks{This work has been funded by projects 6G-CLARION-2022/0005400 and MAP-6G TSI-063000-2021-63, funded by Spanish Ministry of Economic Affairs and Digital Transformation, European Union NextGeneration-EU.}
}
\maketitle
\begin{abstract}
HandOver (HO) control in cellular networks is governed by a set of HO control parameters that are traditionally configured through rule-based heuristics. A key parameter for HO optimization is the Cell Individual Offset (CIO), defined for each pair of neighboring cells and used to bias HO triggering decisions. At network scale, tuning CIOs becomes a tightly coupled problem: small changes can redirect mobility flows across multiple neighbors, and static rules often degrade under non-stationary traffic and mobility. We exploit the pairwise structure of CIOs by formulating HO optimization as a Decentralized Partially Observable Markov Decision Process (Dec-POMDP) on the network’s \emph{dual graph}. In this representation, each agent controls a neighbor-pair CIO and observes Key Performance Indicators (KPIs) aggregated over its local dual-graph neighborhood, enabling scalable decentralized decisions while preserving graph locality. Building on this formulation, we propose \textbf{TD3-D-MA}, a discrete Multi-Agent Reinforcement Learning (MARL) variant of the TD3 algorithm with a shared-parameter Graph Neural Network (GNN) actor operating on the dual graph and \textit{region-wise double critics} for training, improving credit assignment in dense deployments. We evaluate TD3-D-MA in an \texttt{ns-3} system-level simulator configured with real-world network operator parameters across heterogeneous traffic regimes and network topologies. Results show that TD3-D-MA improves network throughput over standard HO heuristics and centralized RL baselines, and generalizes robustly under topology and traffic shifts. 
\end{abstract}


\section{Introduction}
\label{sec:intro}

The densification of cellular networks and the proliferation of heterogeneous deployments significantly increase the complexity of mobility management \cite{tayyab2019surveyho, 3gpp_tr_38874}. As cells become smaller and coverage more irregular, inter-cell interference becomes more severe, while HandOver (HO) decisions become more frequent and more sensitive to channel fluctuations. Meanwhile, modern services ranging from enhanced mobile broadband to Ultra-Reliable Low-Latency Communications (URLLC) impose stringent and often competing requirements on throughput, latency, and reliability. Recent surveys report that legacy Long Term Evolution (LTE)-derived mobility procedures can struggle in ultra-dense and high-mobility regimes, leading to higher radio link failure rates, ping-pong effects, reduced throughput, and load imbalance
\cite{haghrah2023survey5gnr, ullah2023hetnetsurvey, tayyab2019surveyho, ho_ml_survey2021, aiho_lb_survey2024}. These effects are exacerbated by fast variations in radio conditions and heterogeneous user mobility, which increase the sensitivity of HO triggering and parameter settings.

In Third Generation Partnership Project (3GPP) compliant systems, HO behavior is governed by event-based triggers (e.g., A3 events) and a set of HO Control Parameters (HCPs), including hysteresis, Time-To-Trigger (TTT), and Cell Individual Offsets (CIOs), which control HO decisions between neighboring cells. These parameters are traditionally configured offline or adapted through rule-based Self-Organizing Network (SON) functions. Although robust and easy to deploy, such mechanisms are inherently static; a configuration that performs well for one traffic mix and mobility pattern can induce cell-edge starvation, oscillatory HO behavior, or load imbalance under another \cite{ho_ml_survey2021, aiho_lb_survey2024}. This motivates Reinforcement Learning (RL) \cite{SuttonBarto2018} as a data-driven approach for adaptive HO control. A growing body of work formulates HO control as an RL problem either by tuning HCPs or by directly selecting serving cells based on radio measurements \cite{tanveer2022overviewrlho, yajnanarayana2020ho5g_rl, karmakar2022lim2, alsuhli2023mobilityload,  attiah2020loadbalancingrl, chang2023decentralizeddrlmlb}. These studies show that RL controllers can outperform static, heuristic, or rule-based HO baselines in terms of, e.g., throughput, load distribution, or mobility robustness. 

\begin{figure}
    \centering
\includegraphics[width=\linewidth]{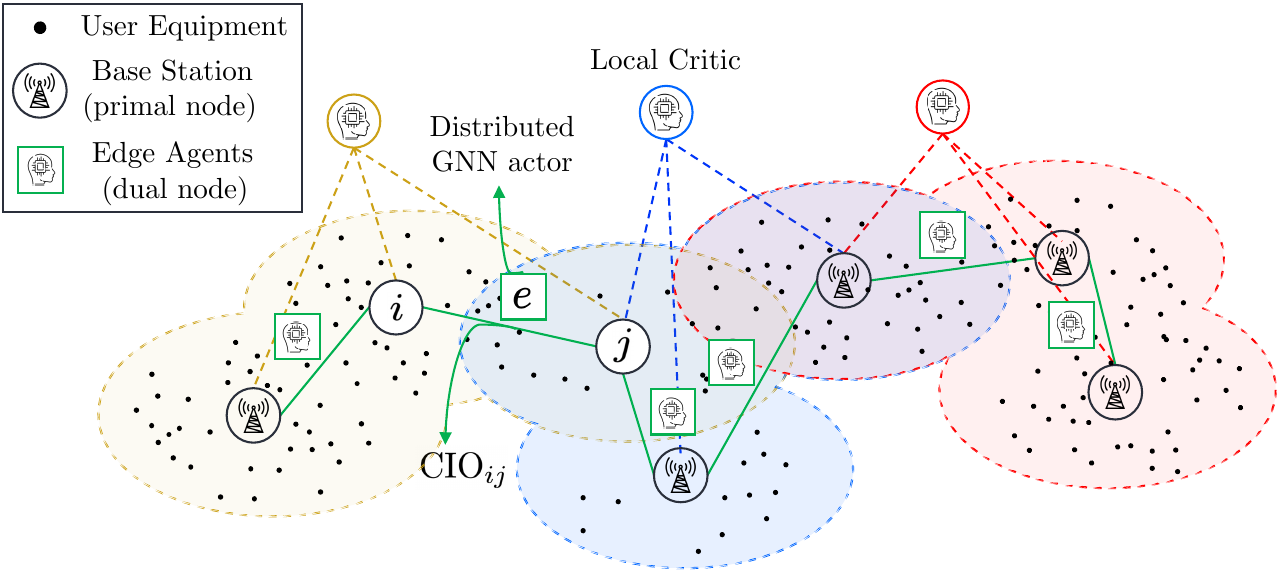}
    \caption{\textbf{Dual-graph MARL framework for CIO-based HO.} CIO agents are placed on dual-graph nodes corresponding to inter-cell edges $e=\{i,j\}$ and tune the HO bias $\mathrm{CIO}_{ij}$ for each neighbor pair. A distributed GNN actor performs local message passing to produce edge actions, while CTDE training uses region-wise critics defined on overlapping primal subnetworks (shaded colored areas).}
    \label{fig:overview}
\end{figure}

\subsection{Motivation, novelty, and contributions} 
Despite this progress, RL-based HO control still faces practical limitations that hinder its scalability, learning stability, and generalization. At network scale, CIO tuning induces a large number of coupled control variables, yielding high-dimensional state representations and multi-discrete action spaces that challenge centralized learning or scenario-specific designs \cite{tanveer2022overviewrlho, alsuhli2023mobilityload}. Moreover, the structure of CIO control is often under-exploited: CIOs are typically treated as a centralized action vector or assigned to per-cell agents, even though CIOs naturally live on \textit{edges} of the network graph, where they induce local yet coupled effects across adjacent cell pairs \cite{alsuhli2023mobilityload, attiah2020loadbalancingrl, chang2023decentralizeddrlmlb}. This coupling also creates credit-assignment problems \cite{SuttonBarto2018}, motivating distributed critic designs that provide more local learning signals in dense deployments \cite{alsuhli2023mobilityload}. While RL methods often show gains in a given scenario, they are seldom evaluated under distribution shift (e.g., changes in topology, traffic, or mobility). Consequently, their generalization behavior is not well understood \cite{tanveer2022overviewrlho, alsuhli2023mobilityload}.
We address the above gaps through the following contributions:
\begin{itemize}
\item We formulate CIO-based HO optimization as a cooperative Multi-Agent Reinforcement Learning (MARL) problem, where each agent controls a single CIO parameter, and agents are placed on CIO edges via a dual-graph representation of the network. Local observations are built from radio cell Key Performance Indicators (KPIs) and aggregated through message passing over the dual graph. To the best of our knowledge, this is the first work to model CIO-based HO control as a MARL problem with agents on CIO edges organized via a dual-graph structure.
\item We introduce TD3-D-MA, a discrete multi-agent extension of TD3 \cite{fujimoto2018td3} with Centralized Training and Decentralized Execution (CTDE). A shared Graph Neural Network (GNN) actor operates on the dual graph, controlling discrete CIO values for each edge agent, while training uses multiple region-wise critics defined on overlapping $N$-hop primal subnetworks to improve credit assignment in dense deployments (see Fig. \ref{fig:overview}).
\item We implement an \texttt{ns-3} \cite{ns3} network simulator configured with real-world network operator parameters, standard A3-triggered HOs, and an RL interface for CIO tuning. In this environment, we benchmark TD3-D-MA against centralized RL and SON-like baselines under heterogeneous network topologies. We compare multiple GNN actor architectures and different critic designs (factorized, non-factorized, local hops input), and show that the dual-graph GNN+CTDE design improves learning stability and generalization across topologies and mobility patterns compared to centralized and rule-based baselines. Our code is available at \href{https://github.com/Telefonica-Scientific-Research/RL4Networks_NS3}{this link}. \end{itemize}

\subsection{Organization}
The remainder of this paper is organized as follows: Section \ref{sec:related} reviews prior RL-based HO and positions this work. Section \ref{sec:problem_formulation} formulates the problem, and Section \ref{sec:algorithm} presents the proposed policy and algorithm. Section \ref{sec:experiments} reports the experimental setting and results, and Section \ref{sec:conclusion} concludes the paper.
\section{Related Work}
\label{sec:related}
Mobility and HO management in wireless networks have been extensively studied from the learning perspective \cite{haghrah2023survey5gnr, ullah2023hetnetsurvey, tayyab2019surveyho, ho_ml_survey2021, aiho_lb_survey2024}. Our work lies at the intersection of $(1)$ RL-based HO optimization; $(2)$ multi-agent and graph-based RL for RAN control; and $(3)$ RL simulation environments for networking. Tab. \ref{tab:related_work_comparison} summarizes representative prior work and highlights how this paper differs along \emph{(a)} \textit{control knob}, \emph{(b)} \textit{learning paradigm}, \emph{(c)} \textit{algorithm}, and \emph{(d)} \textit{simulator}. 

\textbf{RL-based HO optimization.} Most RL-based HO studies adopt centralized or single-agent control and optimize either HO decisions (cell selection) or a subset of HO parameters (see \cite{tanveer2022overviewrlho} for a comprehensive review). Examples include  bandit/RL formulations for mobility decisions \cite{yajnanarayana2020ho5g_rl}, RL-based adaptation of TTT/hysteresis  \cite{karmakar2022lim2}, actor-critic tuning of HO margins \cite{kwong2024ddpgho}, and deep RL for HO decision making in dense HetNets \cite{song2023ho_hetnet_rl}. Similarly to our approach, a few works employ RL to tune CIOs and related offsets/biases, motivated primarily by Mobility Load Balancing (MLB) \cite{rl_lb_survey2023}. Representative approaches include deep RL for CIO-based mobility load management \cite{alsuhli2023mobilityload}, joint CIO and power control \cite{drl_cio_energy_ccnc21}, and RL schemes that optimize offset/neighbor-border parameters in \texttt{ns-3} \cite{attiah2020loadbalancingrl}. Decentralized variants place learning across cells \cite{chang2023decentralizeddrlmlb}, and two-layer schemes combine load prediction with RL-based CIO control \cite{rl_mlb_two_layer_2024}. Recent MARL formulations further tune HO parameters in decentralized settings \cite{shen2023decentralized_marl_ho, elgharably2025hmarl}. These studies demonstrated that adaptive policies can outperform static configurations, but generally remain either centralized/single-agent, focused on a different set of parameters, or evaluated in scenario-specific simulators without reusable environments. 

\textbf{Multi-Agent and Graph-Based RL for HO optimization.} Surveys of MARL for wireless systems highlight challenges such as partial observability, non-stationarity, and scalability, motivating decentralized control and (often) CTDE-style training \cite{marl_wireless_survey2024}. In the mobility management context, decentralized and multi-agent learning has been explored for MLB and HO parameter tuning, e.g., with one agent per cell and actions on HO-related parameters and offsets \cite{chang2023decentralizeddrlmlb, rl_mlb_two_layer_2024, shen2023decentralized_marl_ho, elgharably2025hmarl}. Beyond CIO/parameter tuning, HO-focused MARL also appears in settings such as dense mmWave networks \cite{sana2020marl_mmwave_ho} and joint HO+power or HO+trajectory control \cite{guo2020marlhopower, deng2021uav_trajectory_ho_marl}. Generally, existing MARL formulations for HO/mobility typically place agents on BS (node) entities and act on node-level decisions/parameters, without treating CIO \textit{edges} as first-class decision variables. Our contribution differs by using a dual-graph where CIOs are dual nodes and decentralized actors operate on CIO edges via a shared-parameter GNN under CTDE.

\textbf{RL network simulation environments.}
\texttt{ns-3} is widely used for LTE/NR mobility evaluation \cite{ns3}, and \texttt{ns-3-gym} exposes \texttt{ns-3} as Gym-style environments \cite{gawlowicz2018ns3gym}. Other open-source wireless RL environments exist \cite{wireless_rl_envs}, and there are isolated \texttt{ns-3}-based RL scripts for MRO/HO tuning \cite{ns3_mro_rl_git}, but they generally do not provide a standardized, reusable HO environment focused on 3GPP-style CIO control. Our environment addresses this gap with a CIO-centric interface, multi-topology support, and standardized HO/MLB KPIs.
\begin{table*}[t]
\centering
\caption{Comparison of related work.}
\label{tab:related_work_comparison}
\small
\setlength{\tabcolsep}{4pt} 
\begin{tabular}{@{} l c l l l c @{}}
\toprule
\textbf{Work} & \textbf{Task} & \textbf{RL Type} & \textbf{Simulator} & \textbf{Control Knob} & \textbf{Multi-agent} \\
\midrule

\multicolumn{6}{c}{\textit{HO-focused approaches}} \\
\midrule

\cite{yajnanarayana2020ho5g_rl} &
HO &
Bandit &
5G Simulator &
Beam/Cell Selection &
\xmark \\

\cite{karmakar2022lim2} &
HO &
Single-Agent RL &
\texttt{ns-3} &
TTT + Hysteresis &
\xmark \\

\cite{kwong2024ddpgho} &
HO &
Single-Agent RL &
5G Simulator &
HO Margin &
\xmark \\

\cite{song2023ho_hetnet_rl} &
HO &
Single-Agent RL &
HetNet Simulator &
Cell Selection &
\xmark \\

\cite{sana2020marl_mmwave_ho} &
HO &
MARL &
mmWave Simulator &
Cell Selection &
\cmark \\

\midrule

\multicolumn{6}{c}{\textit{MLB-focused approaches}} \\
\midrule

\cite{alsuhli2023mobilityload} &
MLB &
Single-Agent RL &
LTE Simulator &
CIO &
\xmark \\

\cite{attiah2020loadbalancingrl} &
MLB &
Single-Agent RL &
LTE Simulator &
CIO + Hysteresis &
\xmark \\

\cite{rl_mlb_two_layer_2024} &
MLB &
MARL &
LTE/5G Simulator &
CIO &
\cmark \\

\cite{chang2023decentralizeddrlmlb} &
MLB &
MARL &
Custom Simulator &
Local Bias / HO Offset &
\cmark \\

\cite{mohajer2024madrl} &
MLB &
MARL &
LTE/5G Simulator &
CIO + Bias &
\cmark \\

\cite{shen2023decentralized_marl_ho} &
MLB &
MARL &
5G Simulator &
CIO + Bias &
\cmark \\

\cite{elgharably2025hmarl} &
MLB &
Hierarchical MARL &
LTE / \texttt{ns-3} &
HO Parameters &
\cmark \\

\midrule

\multicolumn{6}{c}{\textit{Combined Control Tasks}} \\
\midrule

\cite{guo2020marlhopower} &
HO+Power &
MARL &
HetNet Simulator &
HO + Power &
\cmark \\

\cite{deng2021uav_trajectory_ho_marl} &
HO+Trajectory &
MARL &
UAV/HetNet Simulator &
HO + Trajectory &
\cmark \\

\midrule

\rowcolor{gray!12}
\textbf{This work} &
\textbf{HO} &
\textbf{CTDE MARL} &
\textbf{\texttt{ns-3}} &
\textbf{Edge CIO} &
\textbf{\cmark} \\

\bottomrule
\end{tabular}

\end{table*}
\section{Problem Formulation}
\label{sec:problem_formulation}
This section formalizes CIO-based HO control as a cooperative MARL problem. We first describe the network model and the HO mechanism, then introduce a Decentralized Partially Observable Markov Decision Process (Dec-POMDP) to model decision making under partial information, and finally instantiate the Dec-POMDP for the CIO control setting.

\subsection{Network Model}
\label{subsec:network_model}
\textbf{Timescales.} We distinguish a \textit{control} timescale and a \textit{measurement} timescale.
Control decisions are taken at discrete epochs indexed by $t=0,1,2,\dots$, separated by
$\Delta>0$ seconds. UE measurements used for HO evaluation (e.g., RSRP reports) are processed
at discrete measurement instants indexed by $n=0,1,2,\dots$, separated by $\Delta_{\mathrm{m}}\ll
\Delta$ seconds. We assume
$\Delta = N_m\Delta_{\mathrm{m}}$, 
for some integer $N_m\in\mathbb{N}$, so that each control interval contains exactly $N_m$ measurement instants. The set of measurement indices belonging to control epoch $t$ is
$ I(t)\triangleq \{tN_m,\,tN_m+1,\dots,(t+1)N_m-1\}. $

\textbf{Cells and network graph.} We consider a mobile network environment with a fixed set of cells $\mathcal{C}= \{1,\dots,C\}$. We represent HO feasibility between network cells through an undirected network graph $\mathcal{G} =(\mathcal{C},\mathcal{E})$, where $\{i,j\} \in \mathcal{E}$ means that cells $i$ and $j$ are \textit{neighbors}, i.e., a UE served by one cell may HO to the other. Neighbor relations are established according to a distance-based heuristic that reflects the spatial proximity of the cells. We denote the \textit{neighbor set} of cell $i$ by $\mathcal{N}(i) \triangleq \{j\in\mathcal{C}:\{i,j\}\in\mathcal{E}\}.$ 

\begin{figure}
\centering
\begin{subfigure}{0.5\linewidth}
    \includegraphics[width=\textwidth]{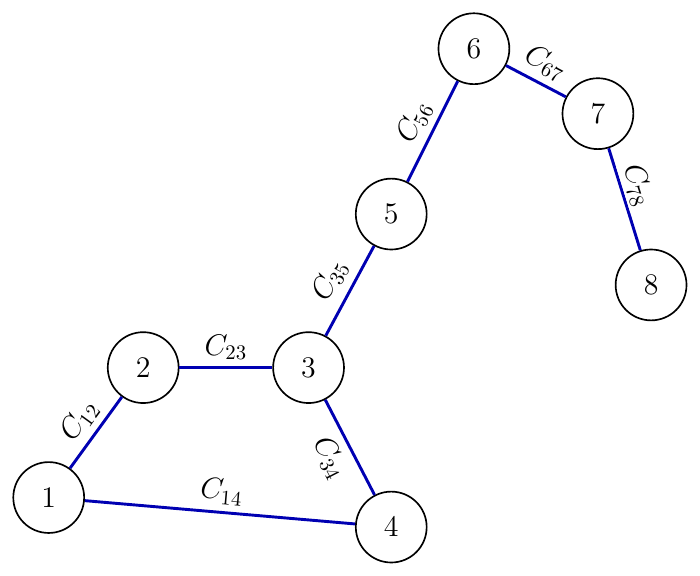}
    \caption{Undirected network graph.}
    \label{fig:graph_representation}
\end{subfigure}
\hfill
\begin{subfigure}{0.43\linewidth}
    \includegraphics[width=\textwidth]{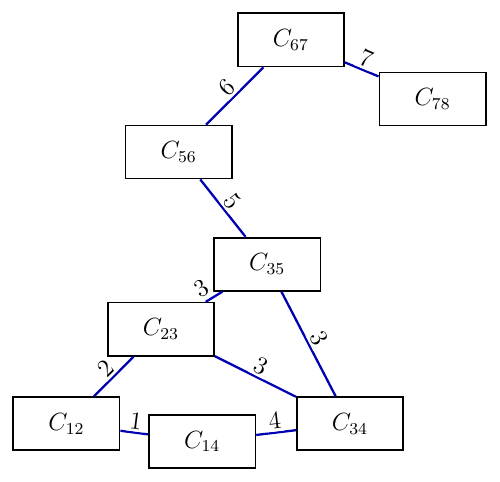}
    \caption{Dual network graph.}
    \label{fig:dual_graph_representation}
\end{subfigure}
\hfill  
\caption{Example of Network graph representations.}
\label{fig:graph_representations}
\end{figure}

\textbf{Dual network graph.} Since CIOs are configured per neighbor edges, it is convenient to define a graph whose nodes correspond to \textit{edges} rather than cells. To this end, we construct a dual graph $\mathcal{G}^\star=(\mathcal{V}^\star,\mathcal{E}^\star)$ from $\mathcal{G}$ as follows:
\begin{itemize}
\item Each dual node corresponds to one undirected neighbor relation (edge) $e=\{i,j\}\in\mathcal{E}$ (controlled CIO variable).
\item Two dual nodes $e$ and $e'$ are adjacent if their underlying edges share at least one endpoint cell, i.e., $e\cap e'\neq\emptyset$.
\end{itemize}
The neighborhood of a node $e\in\mathcal{V}^\star$ in $\mathcal{G}^\star$ will define which  CIO variables are considered ``local'' for decentralized observation aggregation. As the dual-node set and the primal edge set coincide, i.e., $\mathcal{V}^\star \triangleq \mathcal{E}$, we refer to dual nodes simply by $e\in\mathcal{E}$. This dual-graph representation captures the fact that CIOs on adjacent links share cells, and hence their effects on local KPIs are correlated. This observation motivates a decentralized observation model where each agent aggregates information over its dual-graph neighborhood. Note that this construction formally corresponds to the line graph $L(\mathcal{G})$ \cite{wilson1979introduction} (see Fig. \ref{fig:graph_representations} for an example). 

\textbf{Radio quantities.}
Let $\mathcal{U}(n)$ be the set of active UEs at measurement instant $n$, and denote by $\mathcal{U}_i(n)\subseteq \mathcal{U}(n)$ the subset of UEs served by cell $i$ at instant $n$. We define
$ U(n)\triangleq |\mathcal{U}(n)|,$ and 
$U_i(n)\triangleq |\mathcal{U}_i(n)|,$
so that $U(n)=\sum_{i\in\mathcal{C}}U_i(n)$. Each cell $i\in\mathcal{C}$ operates on a fixed system bandwidth $B\;[\mathrm{Hz}]$, partitioned into $N_{\mathrm{PRB}}$ PRBs, such that $B = N_{\mathrm{PRB}}B_{\mathrm{PRB}}$, where $B_{\mathrm{PRB}}\;[\mathrm{Hz}]$ is the PRB bandwidth. At each measurement instant $n$, a scheduler assigns PRBs to connected UEs based on their channel conditions
and traffic demand. At measurement instant $n$ we define the following radio quantities:
\begin{itemize}[leftmargin=*]
    \item $l_u(n)\in\{0,1,\dots,N_{\mathrm{PRB}}\}$: number of PRBs for UE $u$;
    \item $L_i(n)\triangleq \sum_{u\in\mathcal{U}_i(n)} l_u(n)$: total number of PRBs for cell $i$;
    \item $\eta_u(n)\;[\mathrm{bit/s/Hz}]$:  spectral efficiency attained by UE $u$;
    \item $R_u(n)\triangleq l_u(n)\eta_u(n)B_{\mathrm{PRB}}\;[\mathrm{bit/s}]$: downlink rate of UE $u$;
    \item $T_i(n)\triangleq \sum_{u\in\mathcal{U}_i(n)}R_u(n)$:  sum throughput of cell $i$;
    \item $\rho_i(n)\triangleq L_i(n)/N_{\mathrm{PRB}}$: PRB utilization in cell $i$.
\end{itemize}

These quantities are defined at the faster measurement timescale $n$, and capture both resource allocation and link quality effects at each cell and UE. To interface with the controller operating at the slower control timescale $t$, we aggregate the
measurement-scale quantities over the corresponding control interval $I(t)$:
\begin{itemize}[leftmargin=*]
    \item $\rho_i(t)=\tfrac{1}{N_m}\sum_{n\in I(t)}\rho_i(n)$: control-epoch PRB utilization,
    \item $T_i(t)=\tfrac{1}{N_m}\sum_{n\in I(t)}T_i(n)$: control-epoch sum throughput,
    \item $U_i(t)=\tfrac{1}{N_m}\sum_{n\in I(t)}U_i(n)$: control-epoch UE count.
\end{itemize}

We also define a control-epoch channel quality summary $m_i(t)$ (e.g., an MCS/CQI histogram) by aggregating UE reports over $n\in I(t)$, which captures the distribution of channel
qualities experienced within the cell during the control interval.

\textbf{CIO parameters.} CIOs bias the standard HO decision rule and indirectly shape HO directionality and frequency, cell load, and user-level service metrics, such as throughput and blocking probability. While in full generality the offsets are defined on \textit{ordered} serving-target pairs, in this paper we reduce action dimensionality by controlling a single scalar \textit{edge bias} per \textit{undirected} neighbor relation $e=\{i,j\}\in\mathcal{E}$, denoted $b_e(t)$. To represent directed offsets, we fix an orientation for each $e=\{i,j\}$ with $i<j$ and define the antisymmetric mapping
\begin{equation}
\mathrm{CIO}_{ij}(t)\triangleq b_{e}(t),\quad \mathrm{CIO}_{ji}(t)\triangleq -b_{e}(t).
\label{eq:cio_antisym}
\end{equation}
This antisymmetric parameterization preserves HO steering while reducing action dimensionality. For all $n\in I(t)$, the CIO applied at measurement instant $n$ is $\mathrm{CIO}_{ij}(n)\triangleq \mathrm{CIO}_{ij}(t)$, $\forall n\in I(t)$ (and likewise for $(j,i)$).

\textbf{A3 events.} HO conditions are evaluated at the measurement instants $n$. Consider a UE $u$ served
by cell $i$ and a candidate neighbor cell $j\in\mathcal{N}(i)$. Let $\mathrm{RSRP}_{u,i}(n)$ and $\mathrm{RSRP}_{u,j}(n)$ denote
the RSRP measurements available at measurement instant $n$ from cells $i$ and $j$, respectively. Under
the A3 condition (neighbor becomes \textit{``offset better''} than serving), the A3 inequality holds at
instant $n$ iff
\begin{equation}
\mathrm{RSRP}_{u,j}(n)-\mathrm{RSRP}_{u,i}(n) > \mathrm{CIO}_{ij}(n) + H,
\label{eq:a3_ineq}
\end{equation}
where $H>0$ is the hysteresis margin. A handover from $i$ to $j$ is triggered at instant $n$ if
\eqref{eq:a3_ineq} holds for at least $N_{\mathrm{TTT}}\in\mathbb{N}$ consecutive measurement instants,
i.e., if $\forall k\in\{0,\dots,N_{\mathrm{TTT}}-1\}$ it holds that
\begin{equation}
\mathrm{RSRP}_{u,j}(n-k)-\mathrm{RSRP}_{u,i}(n-k)
> \mathrm{CIO}_{ij}(n-k)+H .
\label{eq:a3_rule}
\end{equation}
This provides a discrete-time approximation of the continuous-time TTT mechanism used in LTE/NR when $\Delta_{\mathrm{m}}$ is sufficiently small. Importantly, \eqref{eq:a3_rule} uses the CIO value active at
each instant $n-k$, so the rule remains correct even if the TTT window straddles a control update time. If multiple neighbors satisfy \eqref{eq:a3_rule}, define the eligible set
$ \mathcal{N}_{\mathrm{TTT}}(u,i,n)\triangleq
\left\{ j\in\mathcal{N}(i): \eqref{eq:a3_rule}\ \text{holds for } j \text{ at } n \right\},$
and select the HO target as
$$j^\star(n)\in \operatorname*{arg\,max}_{j \in \mathcal{N}_{\mathrm{TTT}}(u,i,n)}
\big\{\mathrm{RSRP}_{u,j}(n) + \mathrm{CIO}_{ij}(n)\big\}.$$
If $\mathcal{N}_{\mathrm{TTT}}(u,i,n)=\emptyset$, no HO is triggered at instant $n$.

\subsection{RL Formulation}
\label{subsec:rl_formulation}
In the following, we cast the CIO control problem as a cooperative MARL problem using a Dec-POMDP CTDE formulation. We first recall the MDP formalism and then introduce the Dec-POMDP framework used to describe our decision making problem under partial information.

\textbf{Global MDP.} An MDP is defined by a tuple $\mathcal{M}= (\mathcal{S}, \mathcal{A}, P, r, \gamma)$, where $\mathcal{S}$ is the \textit{state space}, $\mathcal{A}$ is the \textit{action space}, $P$ is the \textit{transition kernel}, $r$ is the \textit{reward function,} and $\gamma\in(0,1]$ is the \textit{discount factor}. An agent interacts with the MDP as follows: at each epoch $t$, the agent observes a \textit{state} $s_t \in \mathcal{S}$, selects an \textit{action} $a_t \in \mathcal{A}$, receives a scalar \textit{reward} $r_t = r(s_t, a_t)$, and the system transitions to a new state as $s_{t+1}\sim P(\cdot \mid s_t, a_t)$. As explained next, in our setting decentralization stems from restricting each agent to local KPI information on the dual graph $\mathcal{G}^\star$, while the environment dynamics depends on network-wide configuration.

\textbf{Dec-POMDP.}
A purely centralized policy can become impractical when $(1)$ the number of controllable variables is large (high-dimensional action space); $(2)$ the state aggregates measurements from many entities (high-dimensional state space); and $(3)$ collecting global information at a single decision point induces excessive signaling and latency. These challenges motivate multi-agent control, where multiple cooperative decision-makers each act on a subset of the control variables. Formally, we model our multi-agent setting via a Dec-POMDP defined by $\tilde{\mathcal{M}} = \big( \mathcal{I},\mathcal{S},\{\mathcal{A}^{e}\}_{e \in \mathcal{I}}, P, r,
\{\mathcal{O}^{e}\}_{e \in \mathcal{I}},\Omega,\gamma \big),$ where $\mathcal{I}$ is the set of \textit{agents}; $\mathcal{A}^{e}$ and $\mathcal{O}^{e}$ are the \textit{action} and \textit{observation} spaces of agent $e$; $\mathcal{A} \triangleq \prod_{e \in \mathcal{I}} \mathcal{A}^{e}$ and $\mathcal{O}\triangleq \prod_{e \in \mathcal{I}} \mathcal{O}^{e}$ are the joint action and observation spaces; $P$ is the \textit{transition kernel}; $r:\mathcal{S} \times \mathcal{A} \to \mathbb{R}$ is a shared team reward; $\Omega$ is the \textit{observation kernel}; and $\gamma \in (0,1]$ is the discount factor. For convenience, in the following we identify agents with undirected neighbor edges and write $\mathcal{I} \equiv \mathcal{E}$.

At epoch $t$, the environment is in state $s_t\in\mathcal{S}$. The agents receive local observations $o_t=(o_t^e)_{e \in \mathcal{I}} \in \mathcal{O}$ drawn from an observation distribution $\Omega(\cdot\mid s_t)$, select  actions $a_t^e\sim \pi^e(\cdot\mid o_t^e)$. The joint action $a_t=(a_t^e)_{e \in \mathcal{I}} \in \mathcal{A}$ is applied to the environment, and we observe a global reward $r_t=r(s_t,a_t)$. The objective is to learn decentralized policies ${\pi} = (\pi^e)_{e\in\mathcal{I}}$ maximizing the expected discounted return, defined as
$$ J(\pi)\triangleq \mathbb{E}_{\pi}\!\left[\sum_{t=0}^{T-1}\gamma^t\,r(s_t,a_t)\right].
$$

\textbf{CTDE.} Directly optimizing decentralized policies is challenging because each agent observes only partial information and the effective environment is \textit{non-stationary} from the perspective of individual agents. In CTDE, we train decentralized policies using auxiliary information available only during training. In particular, we learn one or more critics that may condition on additional state-action information (e.g., the global state and joint action, or state-action restrictions and auxiliary local returns defined on local sub-networks, as explained in Sec. \ref{sec:algorithm}), while execution uses only the decentralized policies and local observations $\{\pi^e(\cdot\mid o^e)\}_{e\in\mathcal{I}}$.

\subsection{Dec-POMDP for CIO Control}
\label{subsec:mdp_instantiation}
\textbf{State space.} At each decision epoch $t$, each cell $i\in\mathcal{C}$ measures KPIs
$$ x_i(t) = \big[\rho_i(t),\,T_i(t),\,U_i(t),\,m_i(t)\big], $$
where $\rho_i(t)$ is the PRB utilization, $T_i(t)$ is the downlink throughput, $U_i(t)$ is the number of connected UEs, and $m_i(t)$ summarizes the MCS distribution and assesses the relative channel qualities of UEs in the cell. We assume these KPIs are computed by aggregating measurement-scale quantities over
the indices $I(t)$, and are made
available at each decision epoch. The MCS reflects the channel quality experienced by a UE and is directly determined by the reported Channel Quality Indicators (CQIs), which depend on the received SINR. Lower SINR values result in lower MCS indices, corresponding to more robust modulation and coding. We define the global state as the network-wide KPI feature vector $ s_t = \Big(x_1(t),\dots, x_{C}(t) \Big). $
This KPI-based representation provides a practical abstraction of the network dynamics at the CIO update timescale.

\textbf{Action space.} CIOs are configured per neighbor edge, hence we index one agent per edge, or equivalently, per node of the dual graph $\mathcal{G}^\star$ introduced in Sec. \ref{subsec:network_model}. In particular, $\mathcal{I}\triangleq \mathcal{E}$. Each agent $e=\{i,j\}\in\mathcal{I}$ controls the scalar edge bias $b_e(t)$ and selects it from the discrete set $\mathcal{A}_{\mathrm{CIO}} \triangleq \{-6,-4,-2,0,2,4,6\}\ \text{dB} = \mathcal{A}^{e}$, for all $e\in\mathcal{E}$. 
This discrete set reflects typical bounds and coarse step sizes used in practical SON control, balancing configurability with tractability in multi-agent action spaces.  The joint action space is $\mathcal{A} = \prod_{e\in\mathcal{E}} \mathcal{A}^e = \mathcal{A}_{\mathrm{CIO}}^{|\mathcal{E}|}$, with $a_t = \big(a_t^{e} \big)_{e \in \mathcal{E}},$ and we apply $ b_{e}(t)=a_t^{e}$ until the next decision epoch. 

\textbf{Local observations.} The agents observe locally exchangeable information on the dual graph $\mathcal{G}^\star=(\mathcal{V}^\star,\mathcal{E}^\star)$ with $\mathcal{V}^\star=\mathcal{E}$.
Let $d_{\mathcal{G}^\star}(\cdot,\cdot)$ denote shortest-path distance in $\mathcal{G}^\star$ and define the $M$-hop neighborhood of $e$ as
$\mathcal{N}^\star_M(e)\triangleq \{e'\in\mathcal{E}: d_{\mathcal{G}^\star}(e,e')\le M\}$.
For $e'=\{i,j\}$, define the dual-node feature $z_{e'}(t)\triangleq [x_i(t),x_j(t)]$. The observation of agent $e$ is the attributed rooted subgraph induced by its $M$-hop neighborhood
\begin{equation}
o_t^e \triangleq \Big(\mathcal{G}^\star[\mathcal{N}^\star_M(e)],\ \{z_{e'}(t)\}_{e'\in\mathcal{N}^\star_M(e)}\Big)\in\mathcal{O}^e.
\label{eq:local_obs}
\end{equation}
Intuitively, this observation encodes local KPI summaries for CIO edges within an $M$-hop neighborhood on the dual graph, enabling decentralized agents to condition actions on neighbor network dynamics.

\textbf{Reward.} The reward function $r_t$ maps the state-action pair at epoch $t$ to a scalar objective computed from the above mentioned KPIs. Specifically, we consider the \textit{sum downlink throughput}, i.e., $r_t \triangleq \sum_{i\in\mathcal{C}} T_i(t)$, where $T_i(t)$ is the cell sum throughput (see Sec. \ref{subsec:network_model}). 

While the Dec-POMDP team reward is the global objective $r_t$, under CTDE training we may construct auxiliary \textit{subnetwork returns} $\{r_t^{(j)}\}_{j=1}^J$ by aggregating the same throughput metric over local regions of the
primal graph (as defined in Sec.~\ref{sec:algorithm}). These returns are used only to train region-wise critics for
credit assignment; evaluation remains with the global reward $r_t$.

\textit{Remark.} We also evaluated alternative reward functions including the \textit{percentage of non-blocked UEs} and \textit{average PRB deviation} (load balance), which showed similar trends in our experiments, so we focus on throughput in the main results.
\section{Algorithm}
\label{sec:algorithm}

We present TD3-D-MA, a CTDE multi-agent actor-critic method for the Dec-POMDP setting introduced in Section \ref{subsec:mdp_instantiation}. Agents are CIO edges $e\in\mathcal{E}$ acting on the dual graph $\mathcal{G}^\star$ with local observations $o_t^e$ (see Section \ref{subsec:mdp_instantiation}). The method combines: $(1)$ a discrete TD3-style update using differentiable action relaxations; $(2)$ a shared-parameter GNN actor for decentralized execution; and $(3)$ region-wise double critics trained under CTDE on overlapping $N$-hop primal subnetworks for scalable credit assignment.

\subsection{Critics scope and subnetwork decomposition}
\label{subsec:critic_scope}
We train multiple critics operating on overlapping local regions. While actions correspond to CIO variables indexed by
edges of the neighbor graph, critic inputs are built from \textit{primal} network quantities (cell KPIs), and
therefore critic regions are defined on the primal graph.

\textbf{Primal $N$-hop subnetworks.}
Let $d_{\mathcal{G}}(\cdot,\cdot)$ denote
shortest-path distance on the primal $\mathcal{G}$. For a chosen center node $c^{(j)}\in\mathcal{C}$, define the
$N$-hop neighborhood $\mathcal{N}_N\!\big(c^{(j)}\big)\triangleq \{ i\in\mathcal{C}:\ d_{\mathcal{G}}(i,c^{(j)})\le N\},$
and the induced primal subnetwork
$\mathcal{G}^{(j)} \triangleq \mathcal{G}\!\big[\mathcal{N}_N\!\big(c^{(j)}\big)\big].$ In the distributed configuration, we select $J$ centers $\{c^{(j)}\}_{j=1}^J$ such that the subnetworks
$\{\mathcal{G}^{(j)}\}_{j=1}^J$ jointly cover the deployment (details in Sec.~\ref{subsec:manchester_setup}).
Note that the centralized configuration is recovered as the special case $J=1$ with $\mathcal{N}_N(c^{(1)})=\mathcal{C}$.

\textbf{Induced CIO edge sets.}
Each primal region induces the subset of controllable CIO edges fully contained in the region,
$ \mathcal{E}^{(j)} \triangleq \big\{ e=\{i,k\}\in\mathcal{E}:\ i\in\mathcal{N}_N(c^{(j)}),\ k\in\mathcal{N}_N(c^{(j)}) \big\},$
which specifies the agents and action components visible to critic $j$. (Equivalently, $\mathcal{E}^{(j)}$ is the edge set of the induced subgraph $\mathcal{G}^{(j)}$.)

\textbf{Restriction operator.}
For any global quantity, we denote by $(\cdot)^{(j)}$ its restriction to region $j$. Concretely, $s_t^{(j)}$ collects KPI features only for cells (or sites) in $\mathcal{N}_N\!\big(c^{(j)}\big)$.
Similarly, $a_t^{(j)}$ denotes the sub-vector of the joint CIO configuration that corresponds to the CIO variables indexed by the induced edge set $\mathcal{E}^{(j)}$. Finally, $r_t^{(j)}$ denotes the corresponding subnetwork return obtained by aggregating the same
throughput metric over $\mathcal{N}_N\!\big(c^{(j)}\big)$ (Sec.~\ref{subsec:mdp_instantiation}).

\subsection{Discrete TD3 with categorical relaxations (TD3-D)}
\label{subsec:td3d}

Let $\mathcal{A}_{\mathrm{CIO}}=\{a^{(1)},\dots,a^{(L)}\}$ be the discrete CIO set with $L \triangleq |\mathcal{A}_{\mathrm{CIO}}|$. For each agent $e$, the actor outputs a vector of \textit{logits} (unnormalized log-scores)
$\ell_\theta^e(o_t^e)\in\mathbb{R}^{L}$, and the corresponding categorical distribution $p_\theta^e(o_t^e)\triangleq \mathrm{softmax}\!\big(\ell_\theta^e(o_t^e)\big)$.  

\textbf{Exploration with two mechanisms.}
During environment interaction, each agent uses a two-stage exploration scheme with thresholds
$0 \le \eta_1 \le \eta_2 \le 1$.
At each decision epoch and for each agent $e$, draw $u \sim \mathrm{Unif}(0,1)$ and set
\begin{equation}
a_t^e \triangleq
\begin{cases}
\mathrm{Unif}(\mathcal{A}_{\mathrm{CIO}}), & u < \eta_1,\\
\mathrm{GS}_\tau\!\big(\ell_\theta^e(o_t^e)\big), & \eta_1 \le u < \eta_2,\\
\arg\max_{\ell}\; p_\theta^e(o_t^e)[\ell], & u \ge \eta_2,
\end{cases}
\label{eq:exploration_two_thresholds}
\end{equation}
where $\mathrm{GS}_\tau(\cdot)$ denotes Gumbel-Softmax sampling with temperature $\tau>0$.

A discrete action $a_t^e \in \mathcal{A}_{\mathrm{CIO}}$ is encoded as a \textit{one-hot} vector $\hat a_t^e\triangleq \mathrm{onehot}(a_t^e)\in\{0,1\}^{L}$, while the relaxed (differentiable) action is the probability vector $\tilde a_t^e\triangleq p_\theta^e(o_t^e)$. This relaxed action $\tilde a_t^e$ is used only as a differentiable surrogate for policy optimization, while during the environment interaction the discrete action $a_t^e$ is always executed.

Concatenating over edges yields the joint action embeddings
\begin{equation}
\hat a_t \triangleq [\hat a_t^e]_{e\in\mathcal{E}}\in\{0,1\}^{|\mathcal{E}|L}, \;
\tilde a_\theta(o_t)\triangleq [\tilde a_t^e]_{e\in\mathcal{E}}\in[0,1]^{|\mathcal{E}|L}.
\label{eq:joint_action_embeddings}
\end{equation}
For each region $j$, the critic input uses the restriction to edges in $\mathcal{E}^{(j)}$, denoted by
$\hat a_t^{(j)} \triangleq \hat a_t\!\mid_{\mathcal{E}^{(j)}}$ and
$\tilde a_\theta^{(j)}(o_t)\triangleq \tilde a_\theta(o_t)\!\mid_{\mathcal{E}^{(j)}}$.

\textbf{Action embeddings.} For each region $j\in\{1,\dots,J\}$ we learn two critics $Q_{\psi_1}^{(j)},Q_{\psi_2}^{(j)}$ and corresponding target
critics $Q_{\bar\psi_1}^{(j)},Q_{\bar\psi_2}^{(j)}$. We also maintain an actor $\pi_\theta$ and a target actor with parameters $\bar\theta$. A replay transition is
$(s_t,a_t,r_t,s_{t+1},d_t)$, where $d_t\in\{0,1\}$ is a \textit{terminal flag} ($1$ if the episode ends at $t$, $0$ otherwise). We assume that the local observations $\{o_t^e\}_{e\in\mathcal{E}}$ are deterministically derived from the replayed state $s_t$ via \eqref{eq:local_obs} and the fixed topology $\mathcal{G}^\star$, so they can be reconstructed from replayed states.  


Given this transition, we compute a relaxed next-action from the target actor, $\tilde a_{t+1} \triangleq \tilde a_{\bar\theta}(o_{t+1})$, and its restriction $\tilde a_{t+1}^{(j)}$ for each region.
The clipped double-$Q$ TD target is defined as 
\begin{equation}
y^{(j)} \triangleq r_t^{(j)} + \gamma(1-d_t)\min_{k\in\{1,2\}}Q_{\bar\psi_k}^{(j)}\!\left(s_{t+1}^{(j)},\tilde a_{\bar\theta}^{(j)}(o_{t+1})\right).
\label{eq:td3d_target}
\end{equation}

\textbf{Critics updates.} We train critics on the executed one-hot embedding $\hat a_t^{(j)}$ via the Huber loss  $\mathcal{H}_\delta(x) = \frac{1}{2}x^2 \mathbbm{1}\{|x|\le \delta\} + \delta\left(|x|-\frac{1}{2}\delta\right)\mathbbm{1}\{|x|> \delta\}$, yielding, for $k \in \{1,2\}$,
\begin{equation}
\mathcal{L}(\psi_k^{(j)})\triangleq
\mathbb{E}\Big[\mathcal{H}_\delta\big(Q_{\psi_k}^{(j)}(s_t^{(j)},\hat a_t^{(j)})-y^{(j)}\big)\Big], \; k\in\{1,2\}.
\label{eq:critic_loss_td3dma}
\end{equation}
Note that critics are trained on the executed one-hot actions $\hat a_t^{(j)}$ from the replay buffer, while the TD target and actor update evaluate the critics on relaxed actions $\tilde a^{(j)}$; this continuous relaxation enables stable gradients in the discrete action setting.

\textbf{Actor updates.} Every $d_\pi$ critic updates, we update the actor by maximizing a continuous surrogate objective obtained by evaluating the first critic on the relaxed action. For each region $j\in\{1,\dots,J\}$, we define
\begin{equation}
J^{(j)}(\theta)\triangleq \mathbb{E}\Big[Q_{\psi_1}^{(j)}\!\big(s_t^{(j)},\tilde a_\theta^{(j)}(o_t)\big)\Big],
\;
\theta \leftarrow \theta + \alpha_\theta\nabla_\theta J^{(j)}(\theta),
\label{eq:actor_obj_td3dma}
\end{equation}
and apply this update for all regions $j$ (sequentially, or equivalently by accumulating gradients across regions), 
where $\alpha_\theta>0$ is the actor learning rate.  Target parameters are updated by Polyak averaging with $\tau\in(0,1]$, for $k\in\{1,2\},$
\begin{align}
\bar\psi_k^{(j)} &\leftarrow (1-\tau)\bar\psi_k^{(j)}+\tau\psi_k^{(j)}, \quad \forall j\in\{1,\dots,J\}, \\ 
\bar\theta&\leftarrow (1-\tau)\bar\theta+\tau\theta.
\label{eq:polyak_td3dma}
\end{align}

\subsection{CTDE multi-agent extension (TD3-D-MA)}
\label{subsec:td3dma}
All actors share the same parameters $\theta$ and act based on local  observations only ($M$-hops). Specifically, agents sample actions independently with $\pi_\theta(\cdot\mid o_t^e) = p_\theta^e(o_t^e)$. Exploration during data collection uses the two-stage mechanism in \eqref{eq:exploration_two_thresholds}.

Training and execution follows CTDE: execution is decentralized (each agent uses only $o_t^e$), while critics are available only during training. We decompose the primal graph into overlapping $N$-hop subnetworks $\{\mathcal{G}^{(j)}\}_{j=1}^J$ and train one double critic per region, each conditioning on restricted inputs $(s_t^{(j)},\hat a_t^{(j)})$ and using $r_t^{(j)}$ (Sec.~\ref{subsec:critic_scope}). The centralized critic corresponds to the special case $J=1$ with $\mathcal{N}_N(c^{(1)})=\mathcal{C}$.

\subsection{Function approximation}
\label{subsec:arch_td3dma}

\textbf{GNN actor on $\mathcal{G}^\star$.} The shared actor is a message-passing GNN on the dual graph $\mathcal{G}^\star$. Dual node $e=\{i,j\}$ is initialized with $z_e(t)=[x_i(t),x_j(t)]$ (Sec. \ref{subsec:mdp_instantiation}). After $M$ message-passing layers, each node has an embedding $h_e^{(M)}(t)$ that is mapped by an MLP to logits $\ell_\theta^e (o_t^e) \in \mathbb{R}^{L}$ and hence $p_\theta^e(o_t^e)$. Using $M$ layers implies that the output for node $e$ depends only on $\mathcal{N}_M^\star(e)$, so explicit subgraph extraction is not required.

\textbf{Critic networks.} Each critic maps a state-action pair to a scalar and can be implemented as either $(1)$ an MLP over concatenated features, or $(2)$ a permutation-aware GNN operating on the corresponding primal subnetwork, followed by a readout. Action input is given by $\hat a_t^{(j)}$ (training) or $\tilde a^{(j)}_\theta$ (actor update), with restricted embeddings. The critic receives restricted inputs $(s_t^{(j)},\hat a_t^{(j)})$ (or $(s_t^{(j)},\tilde a^{(j)})$) for each region $j$. We use a non-factorized critic since CIO effects on KPIs are strongly coupled.

\subsection{Training loop}
\label{subsec:training_td3dma}
We store transitions in a replay buffer $\mathcal{D}$ and start gradient updates only after a replay pre-fill of
$N_{\min}$ transitions. We then alternate between environment interaction and off-policy updates from $\mathcal{D}$ with delayed actor updates and Polyak targets (see \eqref{eq:td3d_target}-\eqref{eq:polyak_td3dma}). In the distributed setting, we apply the same update rule in parallel on multiple overlapping $N$-hop subnetworks $\{\mathcal{G}^{(j)}\}_{j=1}^J$ (each with its own critic parameters; replay can be shared with per-region restriction or maintained per critic), and apply the actor update \eqref{eq:actor_obj_td3dma} for each region $j$ (sequentially, or with gradient accumulation).
\begin{algorithm}[!t]
\caption{TD3-D-MA (distributed critics)}
\label{alg:td3dma}
\begin{algorithmic}[1]
\STATE Initialize shared actor $\pi_\theta$ and target actor $\bar\theta\!\leftarrow\!\theta$
\STATE Select critic regions $\{(\mathcal{G}^{(j)},\mathcal{E}^{(j)})\}_{j=1}^J$ (Sec.~\ref{subsec:critic_scope})
\FOR{$j=1,\dots,J$}
  \STATE Initialize double critic $Q_{\psi_1}^{(j)},Q_{\psi_2}^{(j)}$ and targets $Q_{\bar\psi_1}^{(j)},Q_{\bar\psi_2}^{(j)}$
\ENDFOR
\STATE Initialize replay buffer $\mathcal{D}$

\FOR{each episode}
  \FOR{each decision epoch $t$}
    \STATE Run actor on $\mathcal{G}^\star$; sample/apply $a_t$ using \eqref{eq:exploration_two_thresholds}
    \STATE Observe $(r_t,s_{t+1},d_t)$ 
    \STATE Store $(s_t,a_t,r_t,s_{t+1},d_t) \rightarrow \mathcal{D}$
    \IF{$|\mathcal{D}|\ge N_{\min}$}
      \STATE Sample mini-batch; build global $\hat a_t$ and compute global $\tilde a_{\bar\theta}(o_{t+1})$
      \FOR{$j=1,\dots,J$}
        \STATE Restrict batch to region $j$ to obtain $(s_t^{(j)},\hat a_t^{(j)},r_t^{(j)},s_{t+1}^{(j)})$
        \STATE Restrict target action to region $j$: $\tilde a_{\bar\theta}^{(j)}(o_{t+1})$
        \STATE Update critics in region $j$ using \eqref{eq:td3d_target}-\eqref{eq:critic_loss_td3dma}
      \ENDFOR
      \IF{update step $\%\; d_\pi = 0$}
        \FOR{$j=1,\dots,J$}
          \STATE Update actor via \eqref{eq:actor_obj_td3dma} and targets via \eqref{eq:polyak_td3dma}
        \ENDFOR
      \ENDIF
    \ENDIF
  \ENDFOR
\ENDFOR
\end{algorithmic}
\end{algorithm}
\section{Experiments}
\label{sec:experiments}
We evaluate the proposed algorithm on $(1)$ a benchmark environment; and $(2)$ a realistic urban deployment configured using parameters from the Telefónica network in Manchester City (UK). The goal is to study learning behavior and scalability as the HO graph becomes denser, where the number of controllable CIOs is large and credit assignment becomes challenging for centralized training. We first describe the network simulator (Sec. \ref{subsec:sim-impl}) and the baselines (Sec. \ref{subsec:baselines}), the evaluation protocol (Sec. \ref{subsec:eval_protocol}) and then present the results for the benchmark and the realistic Manchester City experiments (Sec. \ref{subsec:network_scenario_benchmark} and Sec. \ref{subsec:manchester_setup}, respectively).

\subsection{Network Simulator and RL interface}
\label{subsec:sim-impl}
All experiments are conducted using \texttt{ns-3} \cite{ns3}, an open-source discrete-event network simulator. The simulator follows a modular architecture that allows the integration of specialized modules that closely follow 3GPP specifications and provide accurate KPIs and a realistic implementation of MAC procedures and 3GPP-compliant HO logic. 

\begin{figure}[h!]
    \centering
    \includegraphics[width=1.04\linewidth]{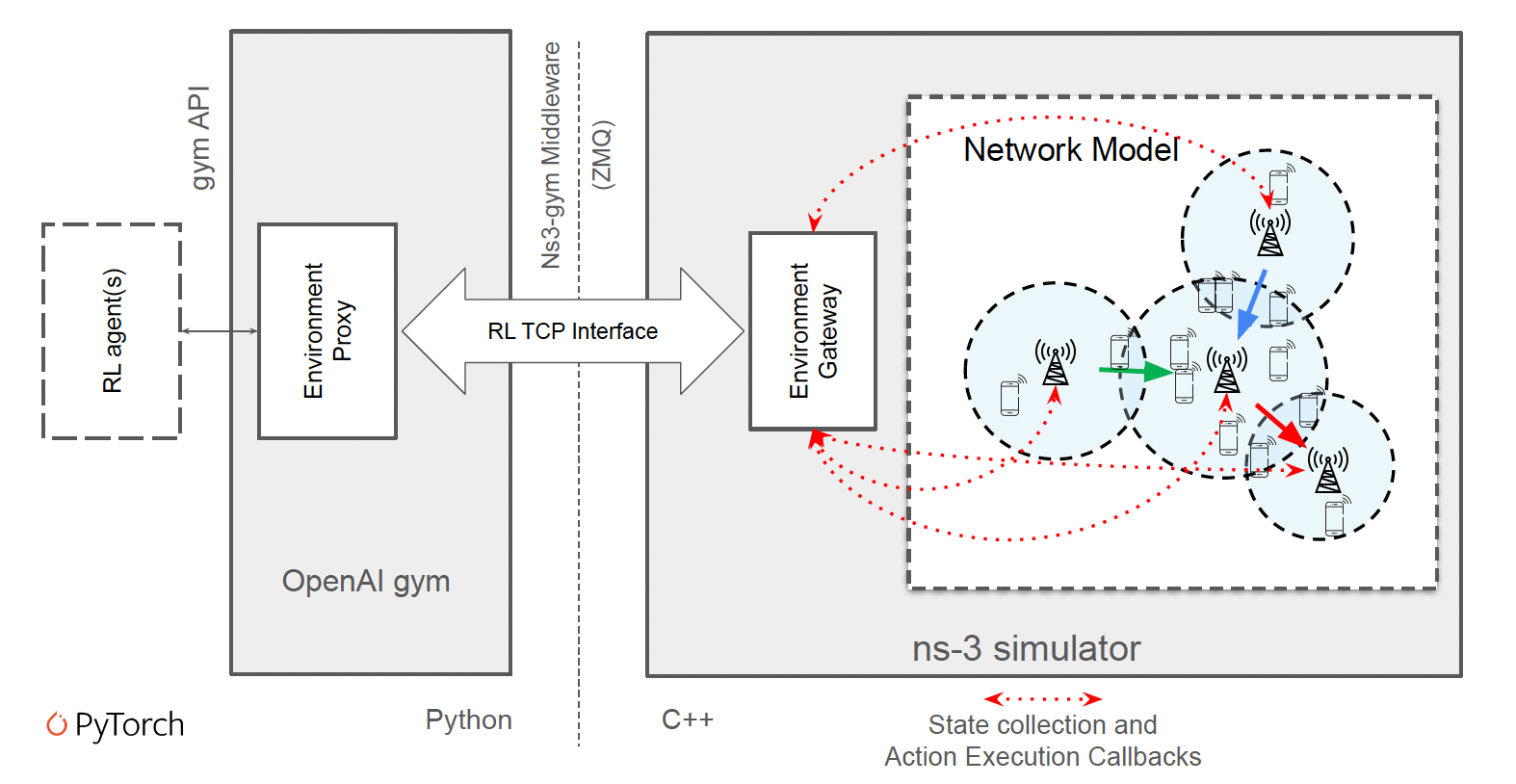}
    \caption{Simulator architecture.}
    \label{fig:architecture_simulator}
\end{figure}

\textbf{Simulation architecture.} Fig. \ref{fig:architecture_simulator} shows the simulator architecture, which is composed of two main elements: the \texttt{ns-3}  etwork simulator and the \texttt{ns-3-gym} \cite{Gawvlowicz19} middleware, which provides the OpenAI Gymnasium interface. The simulator acts as follows: given the current CIO configuration and UE positions, it simulates propagation, scheduling, and HO execution, and returns the next network state and associated KPIs. The \texttt{ns-3} simulator enables the definition of a variety of mobility models, which can be adapted to reflect realistic urban mobility patterns. By configuring these mobility models, we simulate the movement of users within urban areas which would influence network dynamics.

\textbf{RL interface.} To interface the simulator with RL algorithms, we use a Python binding that exposes the \texttt{ns-3} simulator as an environment with the state, action, and reward interfaces. On the algorithmic RL side, we build on Stable-Baselines3 \cite{SB3} and extend it to support parameter-sharing multi-agent RL under CTDE, with decentralized, shared-parameter actors and either a single centralized critic or
multiple region-wise critics defined on overlapping $N$-hop primal subnetworks (Sec.~\ref{sec:algorithm}). Due to the Stable-Baselines3 workflow, we use a single replay buffer that stores a pointer to the sampled topology and
group minibatches by topology when computing critic and actor updates.

\subsection{Baselines}
\label{subsec:baselines}
We compare our method against the following learning-based and heuristic baselines:
\begin{enumerate}[label=$(\roman*)$]
    \item \textbf{Centralized RL baseline (TD3-D-CA):} A single-agent, centralized controller that optimizes all CIOs jointly using global network information as input (single actor and single critic).
    
    \item \textbf{RRM heuristic:} A HO is triggered when the target cell’s RSRP exceeds the serving cell’s RSRP by at least $3 \;\mathrm{dB}$.
    
    \item \textbf{SON heuristic:} A HO is triggered only if the RSRP advantage condition (at least $3 \;\mathrm{dB}$) holds continuously for a fixed TTT of $110 \; \mathrm{ms}$.
    
    \item \textbf{$\Delta$-CIO heuristic:} A rule-based CIO tuning method that adjusts the CIO by a fixed $\Delta = 2 \;\mathrm{dB}$  whenever the load difference between a neighboring cell pair exceeds $15\%$.
\end{enumerate}
\textbf{Critic configuration across scenarios.} Unless stated otherwise, TD3-D-MA uses a centralized critic (the special case $J=1$ in Sec.~\ref{sec:algorithm}) for the benchmark experiments. For the large Manchester scenario, we use the distributed-critic configuration (TD3-D-MA-DC),
with $J>1$ region-wise double critics trained on overlapping $N$-hop primal subnetworks (Sec.~\ref{sec:algorithm}).
All baselines operate at the same control interval and use the same HO hysteresis/TTT configuration unless specified
otherwise. For learning-based baselines, observation and action spaces match TD3-D-MA.

\subsection{Evaluation protocol}
\label{subsec:eval_protocol}
\textbf{Train/test splits.} We consider a set of mobility scenarios partitioned into a training subset and a test subset (Sec.~\ref{subsec:network_scenario_benchmark}). During training, the agent experiences only scenarios from the training subset. We report performance $(1)$ on the training scenarios, and $(2)$ on the held-out test scenarios.

\textbf{Training and evaluation.} Training is performed with exploration enabled (see Sec. \ref{sec:algorithm}), while evaluation is performed with deterministic action selection (greedy action, i.e., $\arg\max$ over the actor logits) and without parameter updates. Results are averaged across multiple random seeds; each episode uses a new seed to randomize UE initial positions and random-walk trajectories, preventing memorization of mobility patterns.

\textbf{Normalization.} Since scenarios differ in UE counts and spatial distributions, raw throughput is not directly comparable across scenarios. For each scenario, we normalize the episode return using scenario-specific lower and upper references: $(1)$ a lower bound given by the best heuristic baseline; and $(2)$ an upper reference obtained by training a centralized single-agent RL controller specialized to that scenario (oracle specialization). The normalized return is
$\bar{r} \triangleq \frac{R - R_{\min}}{R_{\max}-R_{\min}},
$ where $R_{\min}$ is the best-heuristic return and $R_{\max}$ is the oracle specialized-RL return for that scenario.

\subsection{Benchmark environment}
\label{subsec:network_scenario_benchmark}

\textbf{Deployment.} We consider an LTE network composed of $N_c = 8$ cells. The network graph and dual graph are shown in Fig. \ref{fig:graph_representations}. In Fig. \ref{fig:graph_representation}, each node represents an omnidirectional antenna, and edges are labeled with CIO parameters. All antennas use identical radio parameters (carrier frequency, bandwidth, and transmit power). The HO relations define a coupled neighbor graph where only adjacent and selected diagonal cells can exchange users. These relations define $N_e = 8$ controllable CIOs and the dual graph used in the learning architecture.

\begin{figure}[t]
  \centering

  \begin{subfigure}[b]{0.48\linewidth}
    \centering
    \includegraphics[width=\linewidth]{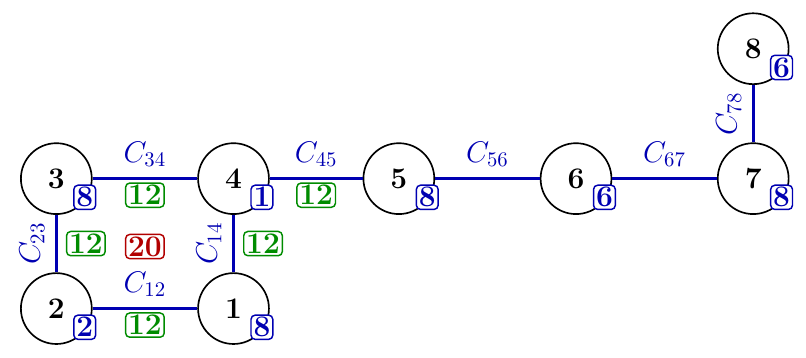}
    \caption{Train 1.}
    \label{fig:train1}
  \end{subfigure}\hfill
  \begin{subfigure}[b]{0.48\linewidth}
    \centering
    \includegraphics[width=\linewidth]{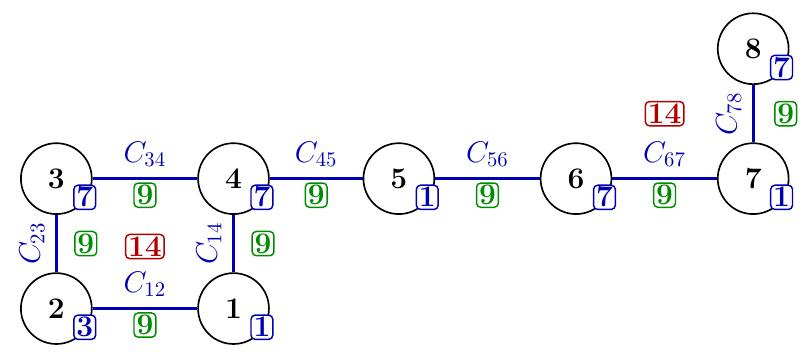}
    \caption{Test 1.}
    \label{fig:test1}
  \end{subfigure}

  \vspace{0.8em}

  \begin{subfigure}[b]{0.48\linewidth}
    \centering
    \includegraphics[width=\linewidth]{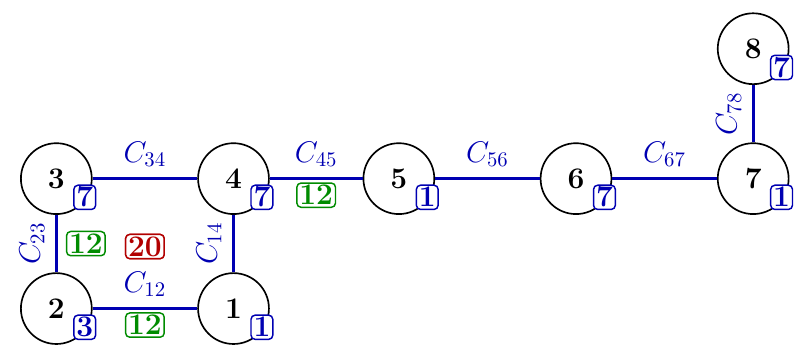}
    \caption{Train 2.}
    \label{fig:train2}
  \end{subfigure}\hfill
  \begin{subfigure}[b]{0.48\linewidth}
    \centering
    \includegraphics[width=\linewidth]{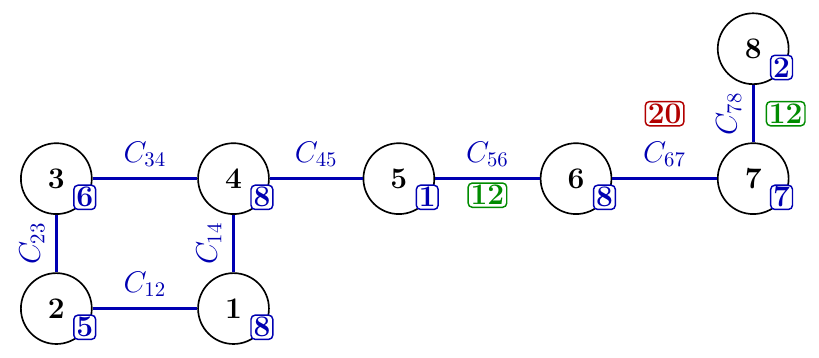}
    \caption{Test 2.}
    \label{fig:test2}
  \end{subfigure}

  \vspace{0.8em}

  \begin{subfigure}[b]{0.48\linewidth}
    \centering
    \includegraphics[width=\linewidth]{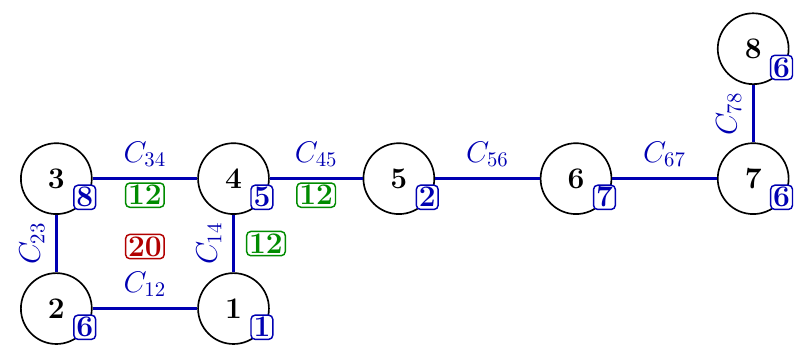}
    \caption{Train 3.}
    \label{fig:train3}
  \end{subfigure}\hfill
  \begin{subfigure}[b]{0.48\linewidth}
    \centering
    \includegraphics[width=\linewidth]{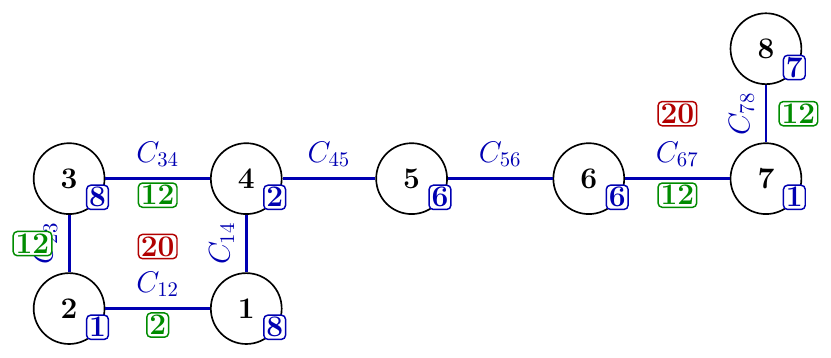}
    \caption{Test 3.}
    \label{fig:test3}
  \end{subfigure}

  \caption{Training (left) and testing (right) mobility scenarios on the 8-cell benchmark. Blue circles denote cells and blue links denote neighbor relations with tunable CIOs. Colored boxes indicate UE random-walk regions: \emph{red} boxes span multi-cell center areas influenced by several neighboring cells, while \emph{green} boxes concentrate mobility near a specific inter-cell border (handover hotspot).}
  \label{fig:mobility_patterns}
\end{figure}

\textbf{Traffic model.}  Each UE generates Constant Bit Rate (CBR) downlink traffic at $1 \;\mathrm{Mbps}$. This ensures that there exists at least one cell that can reach or approach saturation in all mobility scenarios and the network operates under sustained load, making throughput susceptible to HO and load-balancing decisions. Across scenarios, both the UE spatial distribution and the number of active UEs vary, substantially affecting absolute throughput; this motivates reporting normalized return (Sec. \ref{subsec:eval_protocol}).

\textbf{Mobility model.} To study generalization, we construct $6$ mobility scenarios , each defined by a UE spatial distribution and a set of bounded random-walk regions overlaid on the 8-cell topology
(Fig.~\ref{fig:mobility_patterns}). Each UE is assigned to one region type: \textit{red} regions cover multi-cell
center areas where several neighboring cells provide competing candidates, whereas \textit{green} regions localize UEs near a specific inter-cell boundary to stress handover behavior on that neighbor pair. Scenarios differ in the number of UEs placed in each region and in the chosen regions, yielding distinct mobility and load patterns.

\textbf{Matched vs mismatched settings.} We evaluate the models in two regimes: $(1)$ \textit{matched} setting in which training and evaluation are both carried out on the train scenarios. This setting measures how well the agent can adapt to stochastic variations within a fixed set of mobility patterns; $(2)$ \textit{mismatched} setting where the agent is trained only on the train scenarios and evaluated on test scenarios which were not observed during training. This setting quantifies the ability of the learned policy to generalize to unseen UE distributions and trajectories.

\begin{figure}[!htb]
  \centering

  \includegraphics[width=0.75\linewidth]{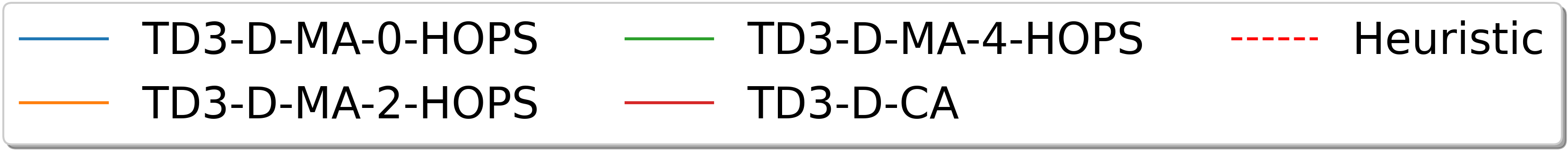}\par\vspace{2mm}

  \begin{subfigure}[t]{0.49\linewidth}
    \centering
    \includegraphics[width=\linewidth]{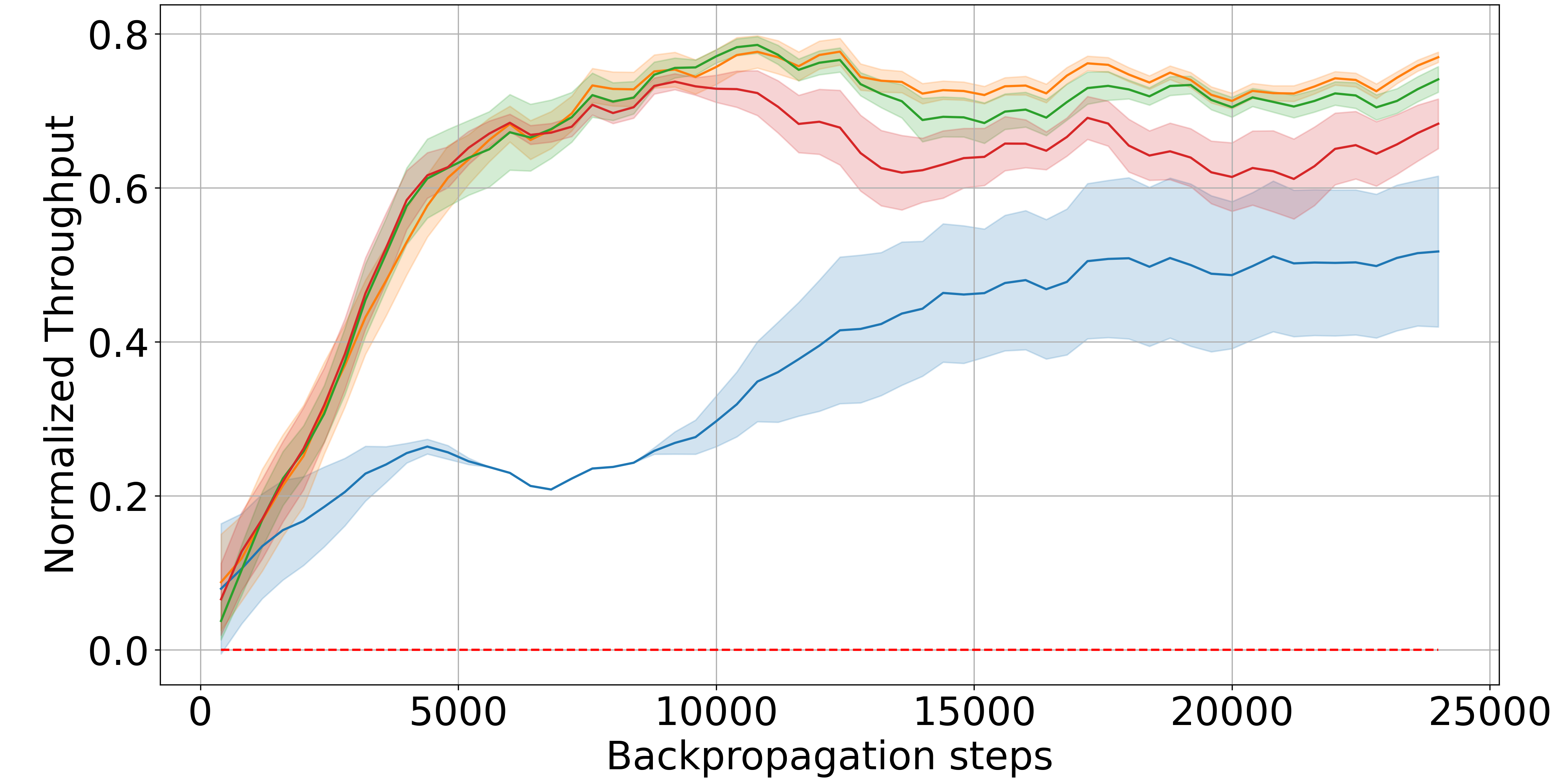}
    \caption{Matched setting.}
  \end{subfigure}\hfill
  \begin{subfigure}[t]{0.49\linewidth}
    \centering
    \includegraphics[width=\linewidth]{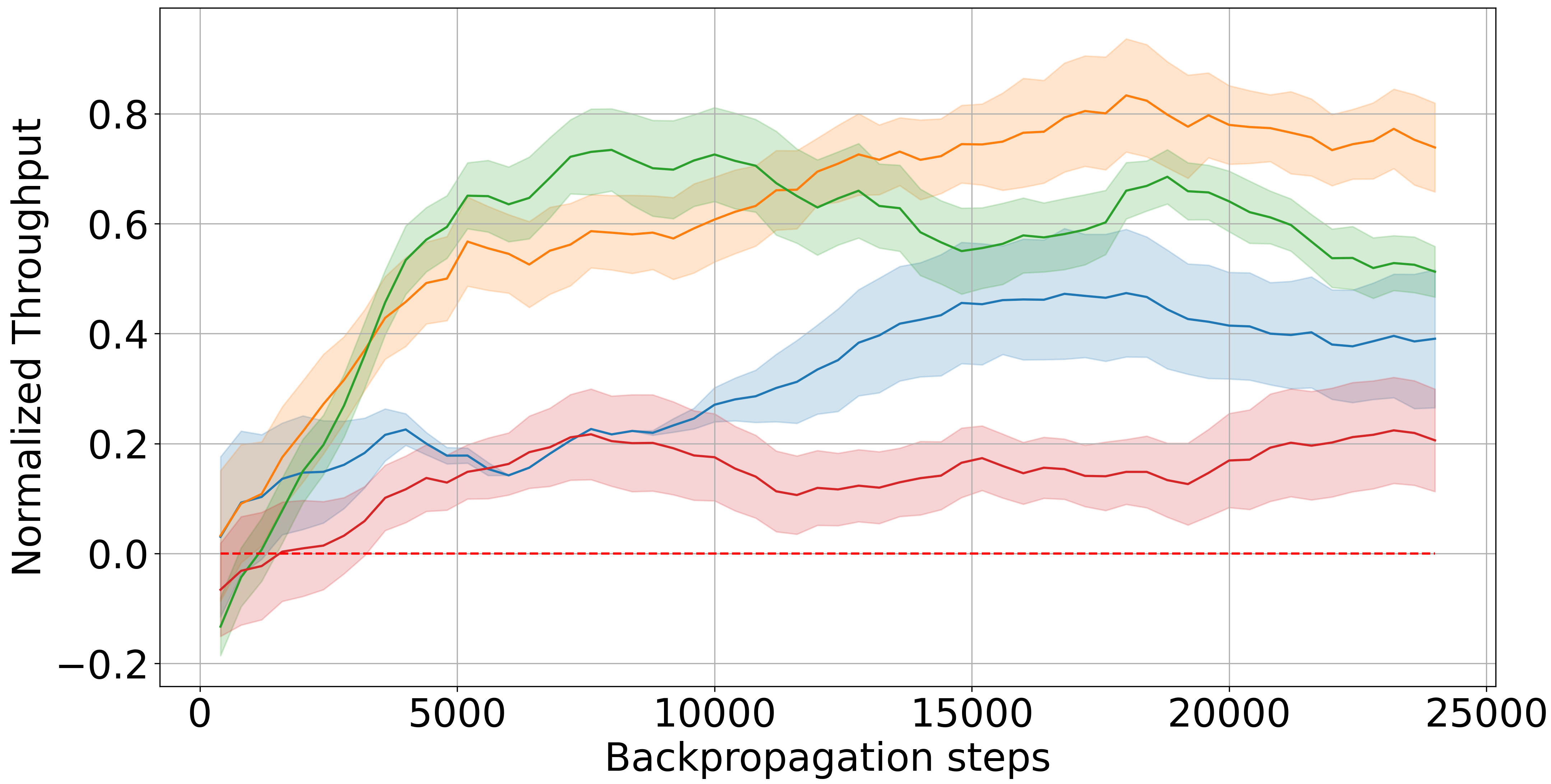}
    \caption{Mismatched setting.}
  \end{subfigure}

  \caption{Impact of the number of hops ($M$) on performance.}
  \label{fig:number_hops_comparison}
\end{figure}

\textbf{Impact of neighbors.} The number of hops included in the actor’s input greatly impacts performance. More information provides the actor with a comprehensive understanding of the network status, enabling more robust decision-making. However, larger neighborhoods require collecting and sharing more KPI information, increasing signaling overhead and
latency in practical implementations. Therefore, it is crucial to understand how performance is affected by the number of hops. The results of this analysis are shown in Fig. \ref{fig:number_hops_comparison} for both the matched (left) and mismatched (right) settings. 

Regardless of the number of hops, the multi-agent approach generally performs better than the centralized one. Using $2$-hops provides enough information to effectively solve the problem. Employing $4$-hops leads to faster convergence (approximately $11,000$ vs. $16,000$ training steps) but requires transferring more information across the network. 

\begin{figure}[!htb]
    \centering
    \includegraphics[width=1\linewidth]{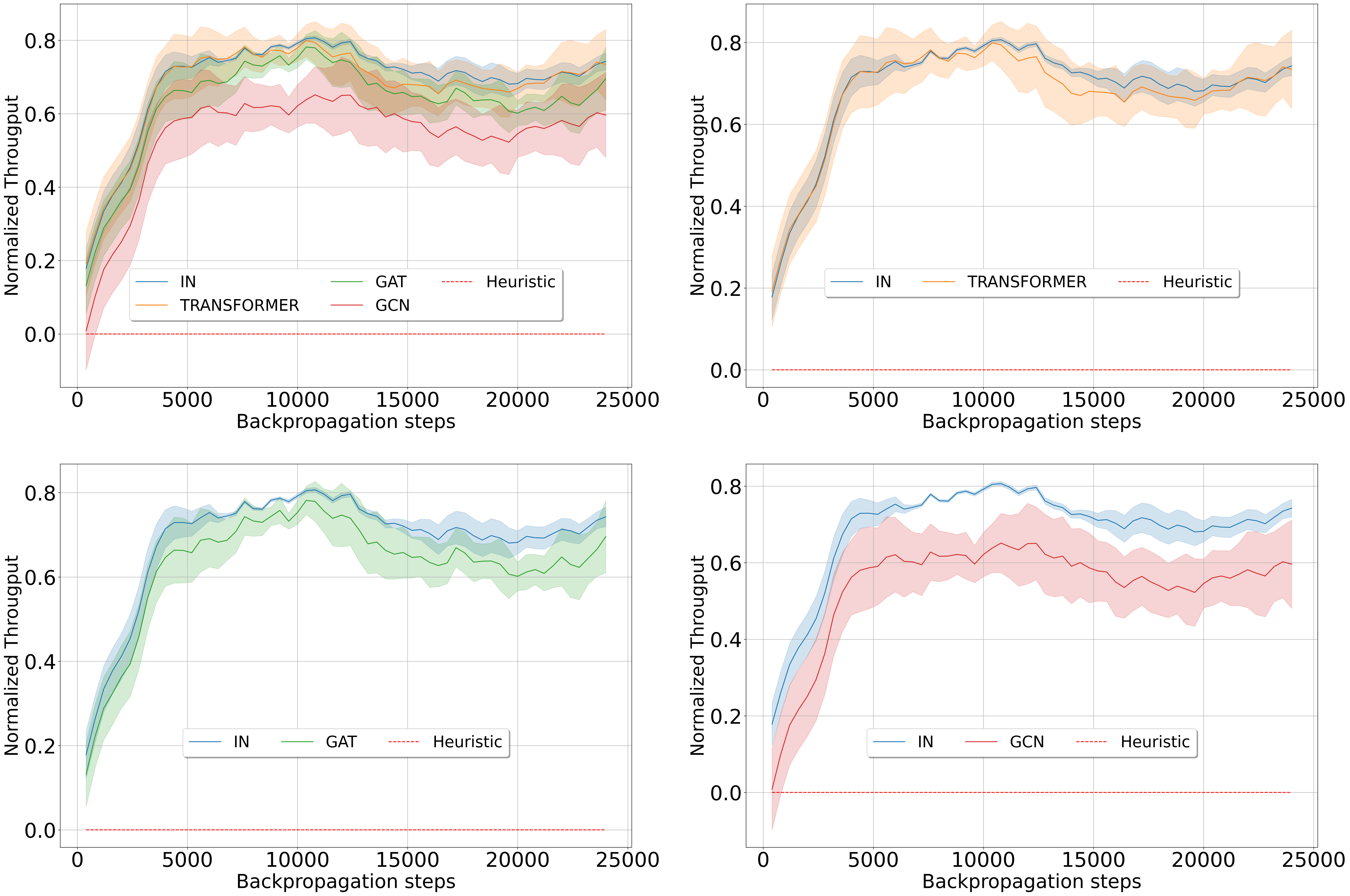}
    \caption{Impact of the Actor GNN on model performance. In all cases, the actor receives information from $2$-hop neighbors.}
    \label{fig:comparison_gnns}
\end{figure}

\textbf{Actor architecture ablation.}
We next study how the GNN design impacts performance. We compare a Graph Convolutional Network (GCN) \cite{kipf2017semisupervised}, Graph Attention Network (GAT)
\cite{velivckovic2018graph}, Transformer \cite{vaswani2017attention}, and Interaction Network (IN)
\cite{battaglia2016interaction, battaglia2018relational}. In all cases, the actor uses the same observation radius
($2$-hop neighbors) and differs only in the message-passing architecture. Fig.~\ref{fig:comparison_gnns} reports normalized return during training for the matched and mismatched settings.

From the results, we conclude that $(1)$ GAT, IN and Transformer architectures consistently outperform GCN; $(2)$ both transformer and IN outperform GAT; $(3)$ IN demonstrates greater stability and exhibits lower variance compared to transformer. This suggests that, while GCN utilizes graph topology without discerning the importance of existing edges, GAT, Transformer, and IN employ soft-link-prune mechanisms. Moreover, these mechanisms allow them to learn graph edge weights, distinguishing the contribution of neighboring nodes to contextual node updates. The progression from GAT to Transformer to IN reflects an increase in the complexity of the mechanisms for learning edge weights. In our problem setting, where distinguishing the state of neighboring cells is critical for  performance, this progression explains the observed improvement in performance.

\begin{figure}[!htb]
    \centering

    \includegraphics[width=0.95\linewidth]{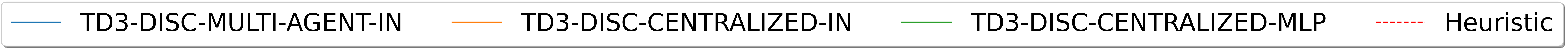}\par\vspace{0.5em}

    \begin{subfigure}[t]{0.49\linewidth}
        \centering
        \includegraphics[width=\linewidth]{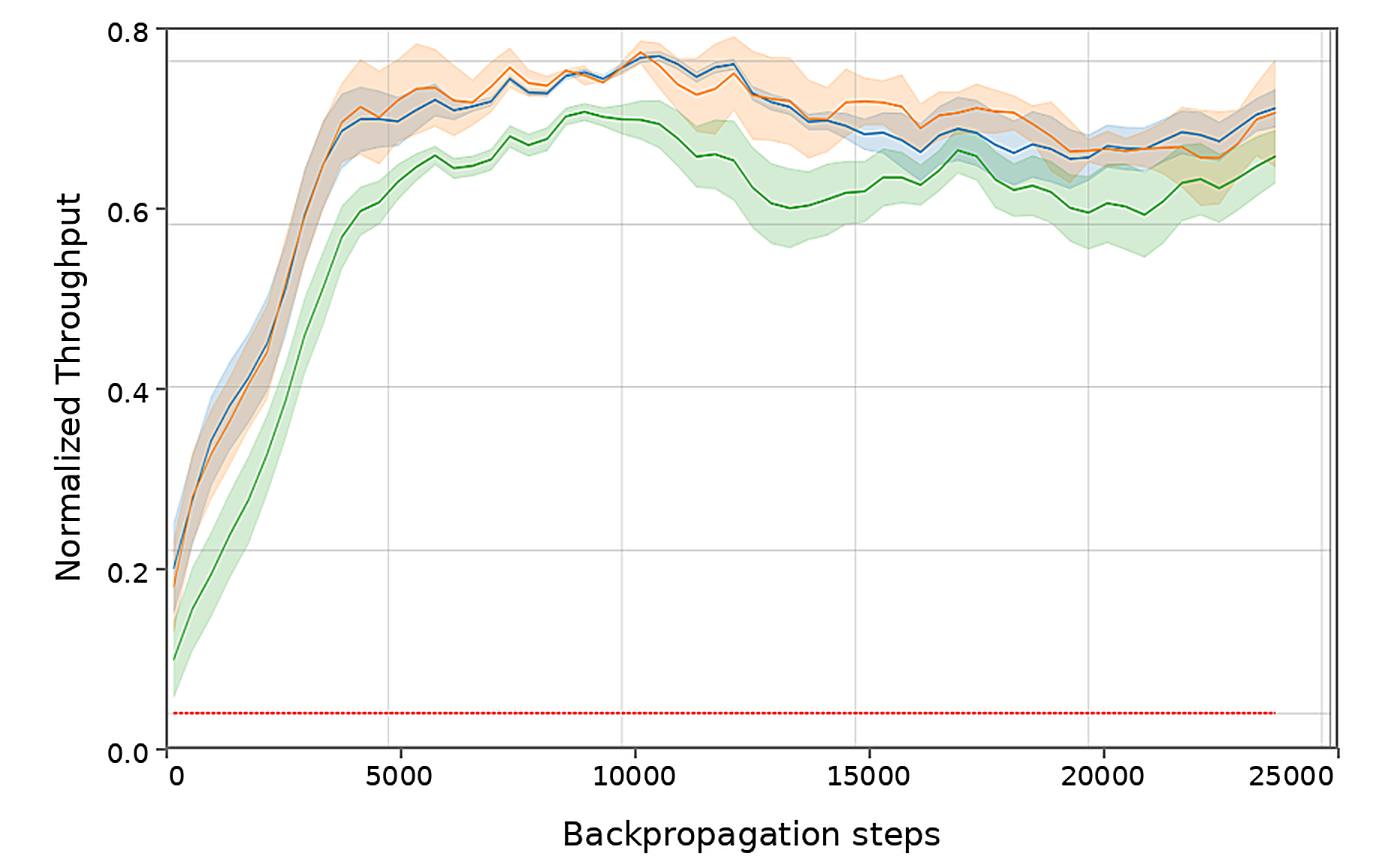}
        \caption{Training scenarios.}
        \label{fig:matched}
    \end{subfigure}\hfill
    \begin{subfigure}[t]{0.49\linewidth}
        \centering
        \includegraphics[width=\linewidth]{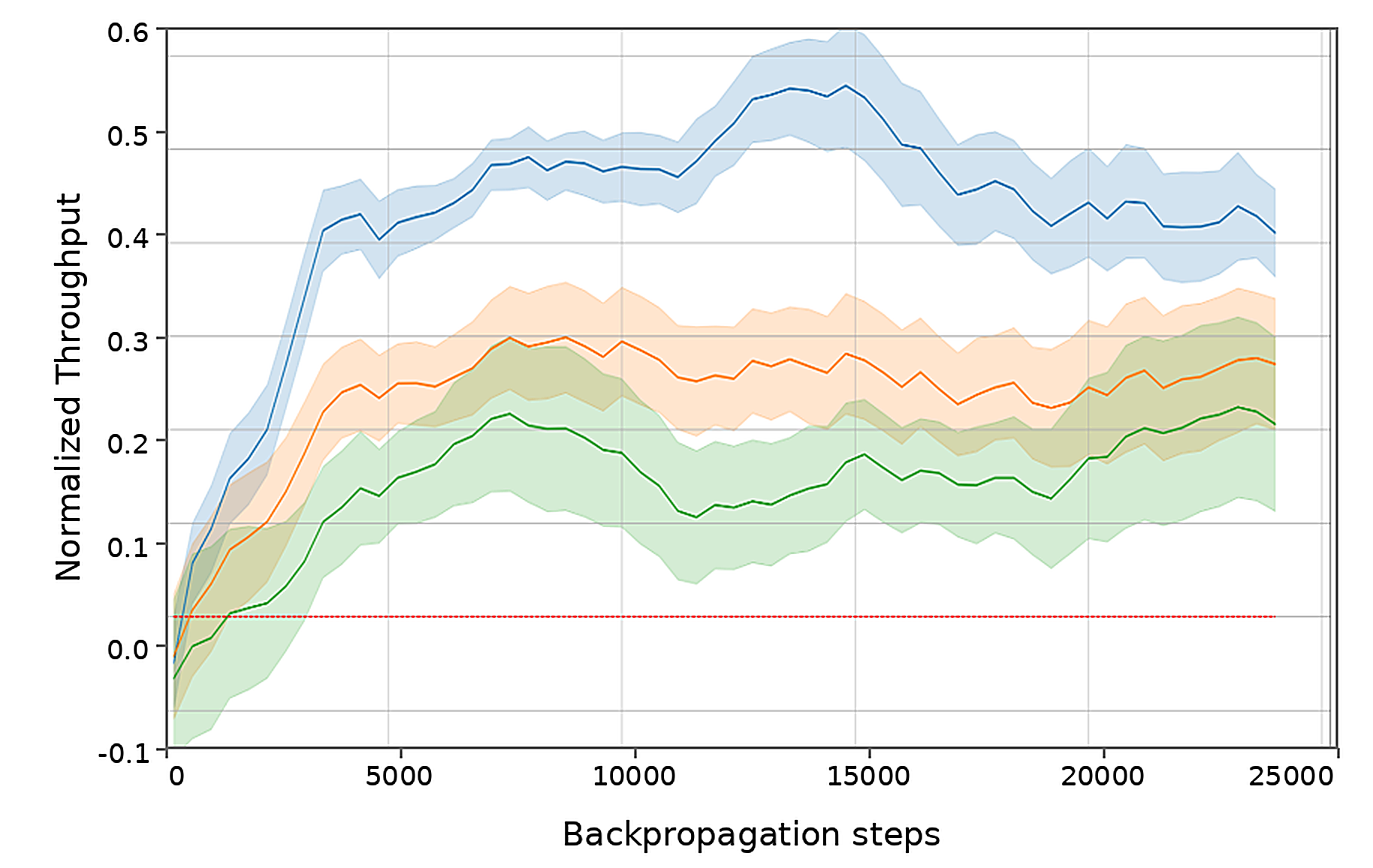}
        \caption{Test scenarios.}
        \label{fig:mismatched}
    \end{subfigure}

    \caption{Generalization capability: centralized-MLP-actor vs. centralized-GNN-actor vs. distributed-GNN-actor.}
    \label{fig:generalization}
\end{figure}

\textbf{Generalization under topology shifts.} Next, we evaluate the contribution of the GNN embedding and the distributed (dual-graph) actor. We compare three models:
$(1)$ \textit{centralized-MLP-actor}: the centralized baseline of  Sec.~\ref{sec:algorithm} with an MLP actor;
$(2)$ \textit{centralized-GNN-actor}: a centralized actor where the first embedding is replaced by a GNN (IN), but the
readout remains an MLP; and $(3)$ \textit{distributed-GNN-actor}: our dual-graph GNN actor with parameter sharing.
Fig.~\ref{fig:generalization} reports performance on training scenarios (left) and test scenarios (right). For fairness, all GNN-based actors use the IN backbone identified in Fig.~\ref{fig:comparison_gnns}.

Comparing the two centralized actors show that using a GNN to enhance node representation improves the overall
model performance. Specifically, in test scenarios, the centralized GNN actor demonstrates superior generalization compared to the centralized MLP one. From the results we can conclude the following: $(1)$ Both the use of GNNs and the distributed approach enhance generalization capabilities; $(2)$ The generalization capability of the centralized-GNN-actor is limited by the readout MLP, which lacks permutation-equivariant structure at the readout stage, which can hinder generalization across topology variations.

\begin{figure}[!htb]
    \centering
    \includegraphics[width=\linewidth]{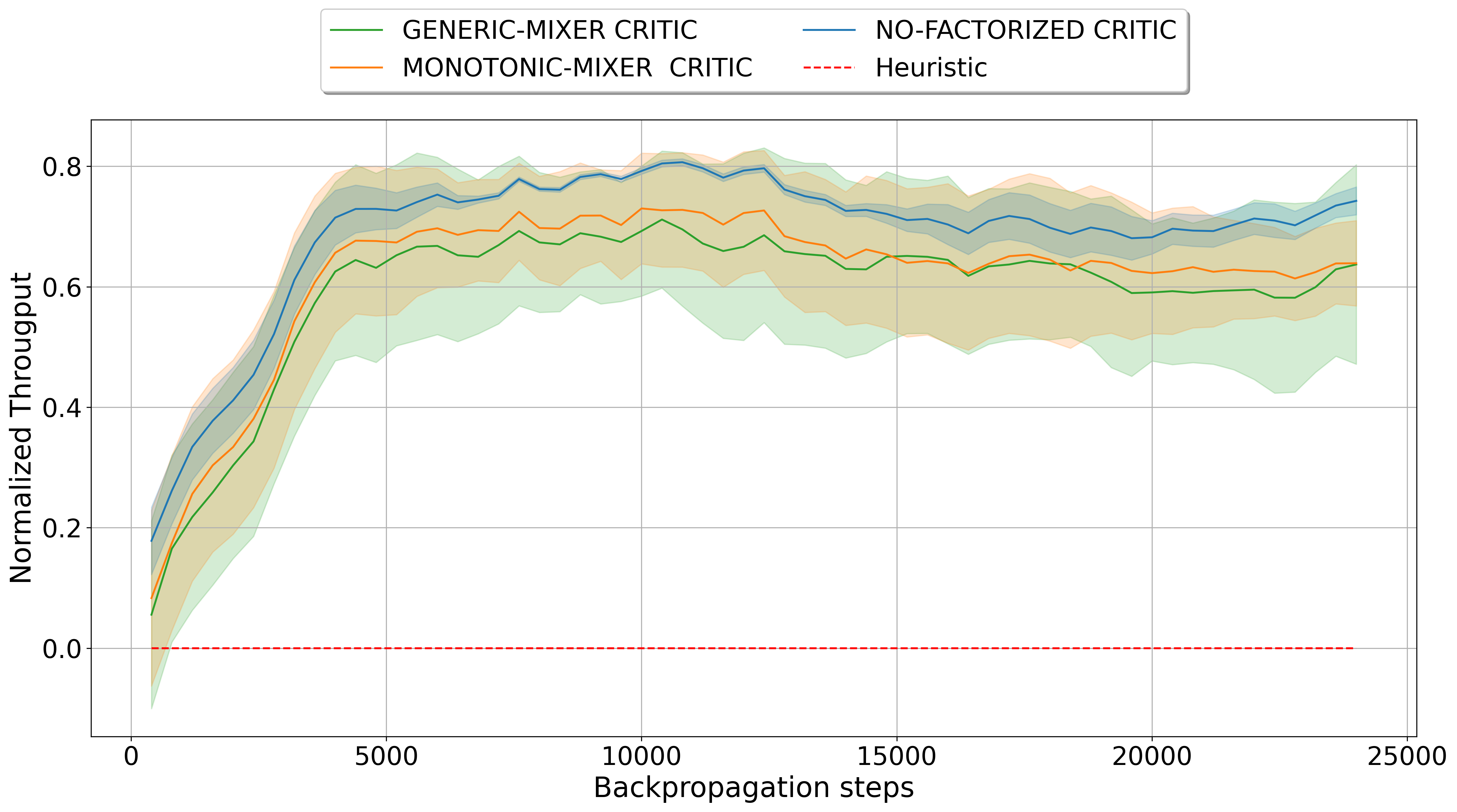}
    \caption{Impact of different type of factorizations for the critic.}
    \label{fig:critic_design_impact}
\end{figure}

\textbf{Critic architecture.} In the 8-cell benchmark, the critic is implemented as a centralized MLP over concatenated KPI features and the joint action embedding. We evaluate the impact of the critic architecture on the performance of the model. We compare model performance using different critic implementations: non-factorized critic, monotonic mixer critic (QMIX) \cite{rashid2018qmix}, and generic mixer critic. All critic variants in this ablation use the same (centralized) critic input for the benchmark setting; only the
critic parameterization differs (non-factorized vs.\ mixing). The results are shown in Fig. \ref{fig:critic_design_impact}. Regardless of the implementation, the factorized critic consistently yields inferior performance compared to the non-factorized critic. This is primarily due to the high variance observed during the learning process. 

\subsection{Manchester Network scenario}
\label{subsec:manchester_setup}

\begin{figure}[!htb]
    \centering
    \begin{subfigure}[t]{0.495\linewidth}
        \centering
        \includegraphics[height=4cm]{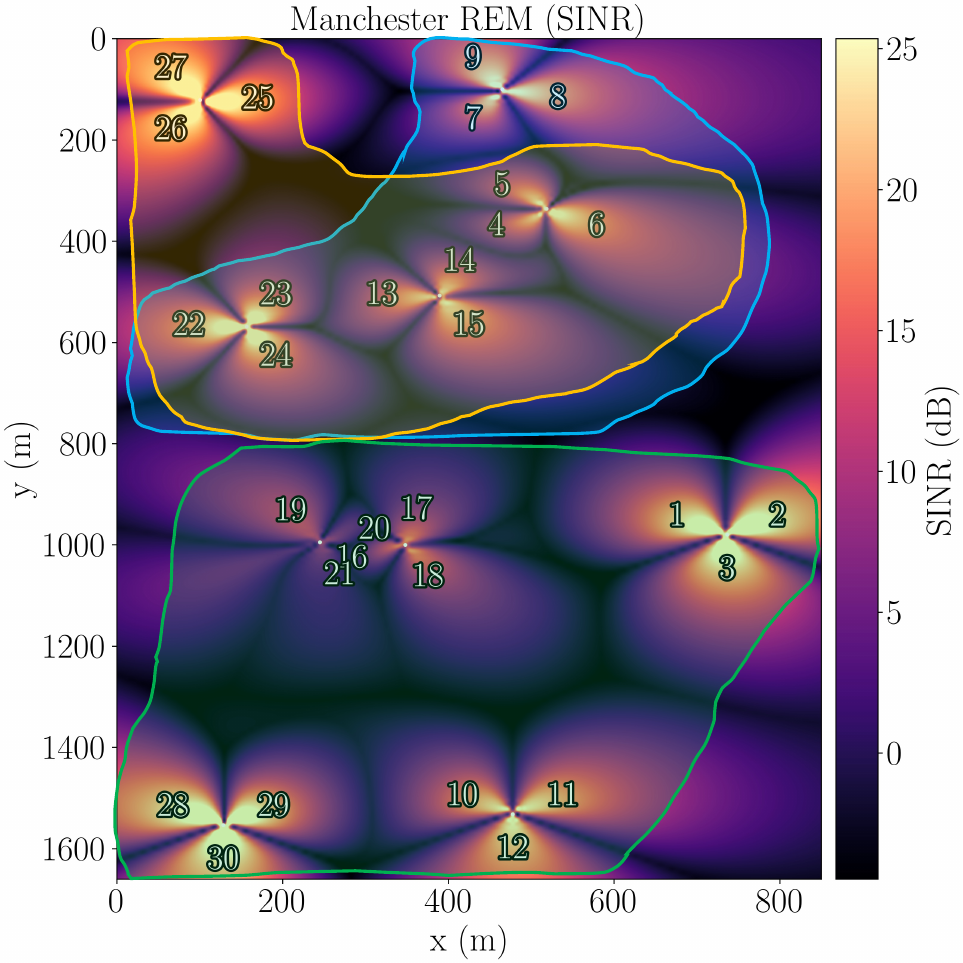}
        \caption{REM for sub-networks North-$1$ (yellow), North-$2$ (blue), and South (green).}
        \label{fig:manchester_rem}
    \end{subfigure}\hfill
    \begin{subfigure}[t]{0.495\linewidth}
        \centering
        \raisebox{0.2\height}{%
            \includegraphics[height=3cm]{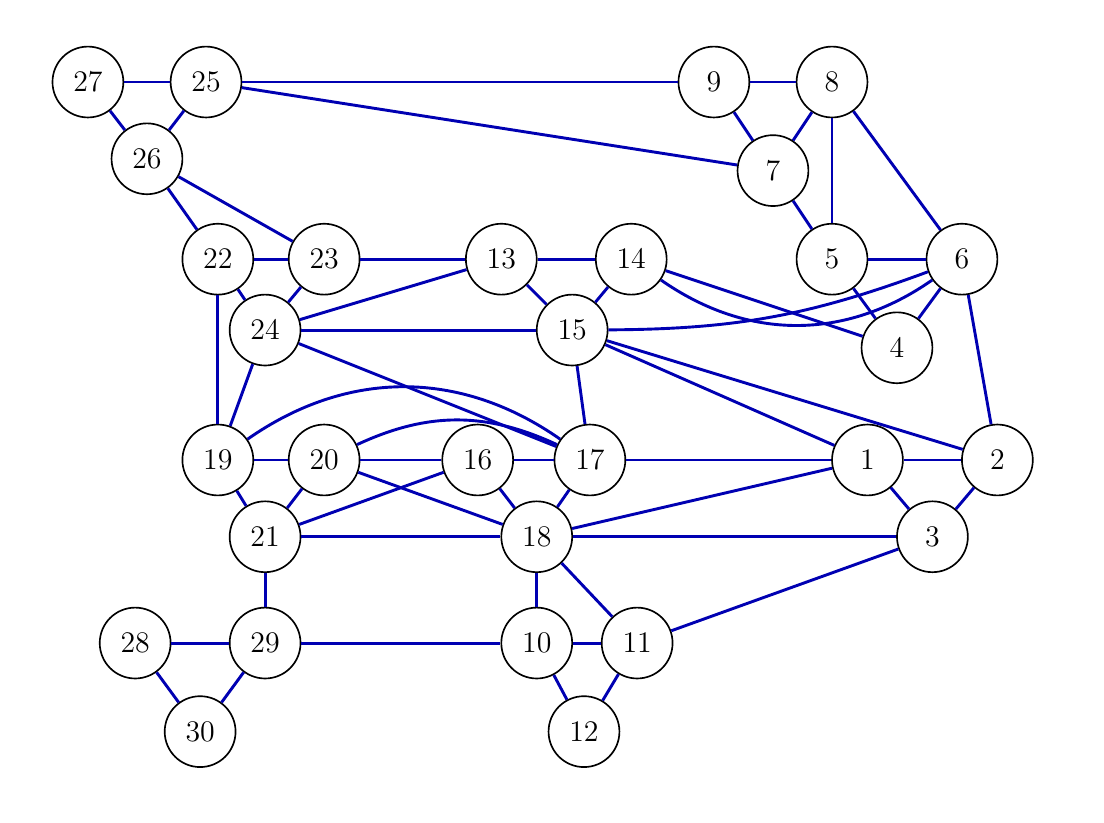}
        }
        \caption{Corresponding network graph for the Manchester scenario.}
        \label{fig:scheme_antennas_manchester}
    \end{subfigure}

    \caption{Manchester scenario: radio environment map and corresponding network graph.}
    \label{fig:manchester_rem_and_graph}
\end{figure}

\textbf{Deployment.} We consider a network area of approximately $2~\mathrm{km}^2$ around Albert Square in Manchester City, comprising $10$ sites with $3$ directional antennas per site ($N_c = 30$). Antenna locations and radio parameters (transmit power, carrier frequency, bandwidth, azimuth, height) are configured to match the operator setup from the real-world network deployment. Fig. \ref{fig:manchester_rem} shows the Radio Environment Map (REM) for the SINR [dB] in the Manchester City scenario. The graph has an average graph degree of approximately $4.3$, and a number of CIO parameters (local action space dimensionality) of $64$.
In this regime, a centralized Critic that learns from a single aggregated return (e.g., network throughput) must attribute outcomes to a large set of simultaneous actions, which makes credit assignment harder and can slow or destabilize learning. This motivates revisiting the Critic design.

\textbf{Users and traffic.}
The scenario includes  $N_{\text{UE}} = 300$ UEs, and a heterogeneous mix of downlink traffic profiles: $(1)$ full-buffer traffic with MBR $=20 \;\mathrm{Mbit/s}$  (file transfer/cloud sync); $(2)$ bursty $3 \;\mathrm{Mbit/s}$ (video streaming); $(3)$ bursty $750 \;\mathrm{Kbit/s}$  (web browsing); $(4)$ bursty $150 \;\mathrm{Kbit/s}$ with guaranteed bit rate (voice calls). 

\textbf{Experimental configurations.} All experiments in this section use the same distributed GNN-based actor described in Sec.~\ref{sec:algorithm}. We evaluate multiple critic configurations while keeping the actor, observation space, action space, and training hyperparameters fixed, to isolate the effect of credit assignment in dense graphs. In the distributed-critic variant (TD3-D-MA-DC), we instantiate one \emph{GNN-based} double critic per region, where each critic operates on the corresponding primal subnetwork $\mathcal{G}^{(j)}$ and produces a scalar value via a graph-level readout (see Sec.~\ref{sec:algorithm}). 

\begin{figure}[!htb]
    \centering
        \includegraphics[width=0.6\linewidth]{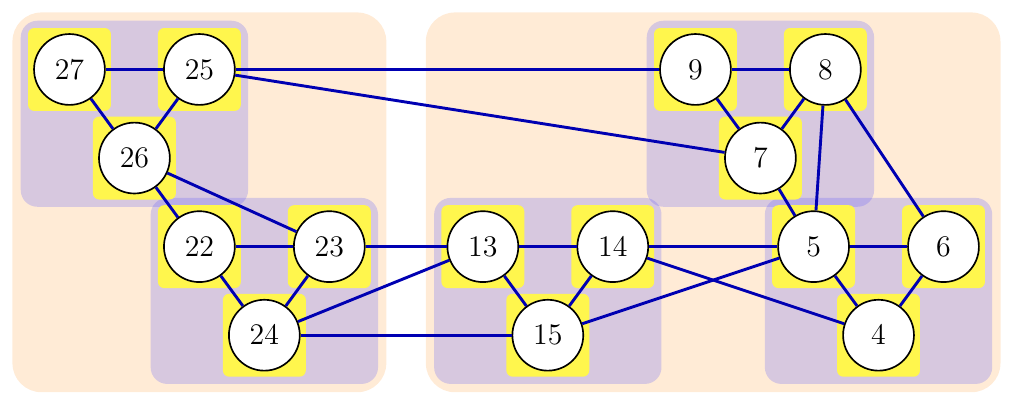}
    \caption{Reward decomposition for CTDE training: global return, and local returns over overlapping $N$-hop subnetworks (shaded boxes).}
    \label{fig:experimental_config}
\end{figure}

\begin{figure}[!htb]
    \centering
    \includegraphics[width=0.75\linewidth]{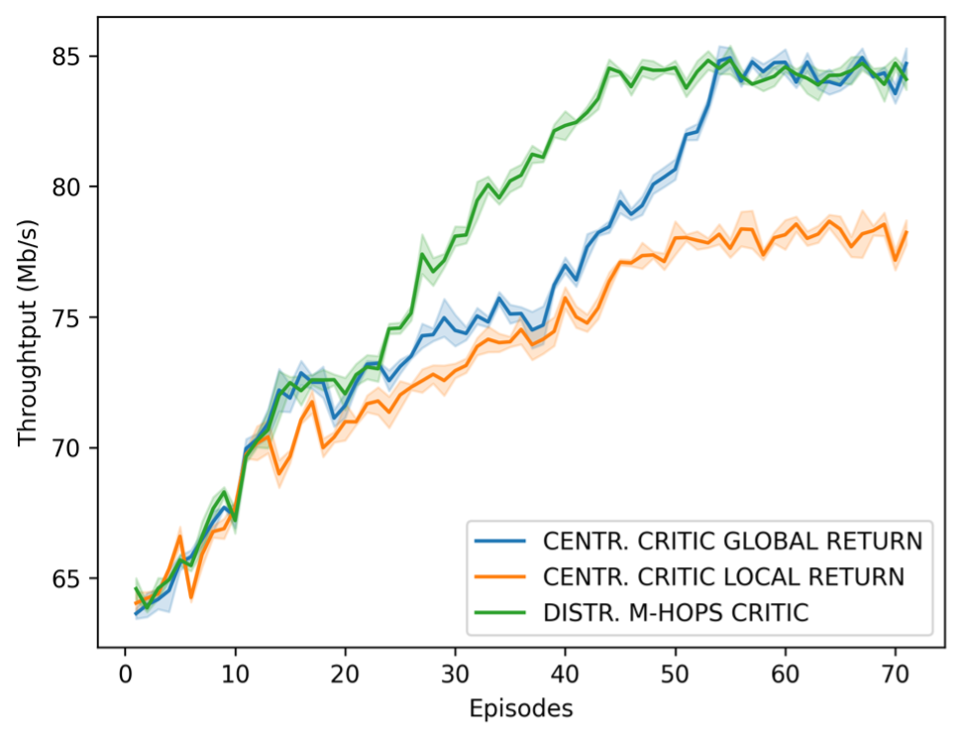}
    \caption{Critics comparison results.}
    \label{fig:critics_comparison}
\end{figure}

Fig. \ref{fig:experimental_config} illustrates different locality levels used when defining local returns in centralized training. These choices influence the granularity of the learning signal and help explain the performance differences observed in Fig. \ref{fig:critics_comparison}. Note that these locality levels modify only the critic's training target (credit assignment) under CTDE, while the underlying environment objective remains the same team-level reward. For the distributed critic configuration, the Manchester graph is decomposed into $J=4$ overlapping sub-networks using $N=2$ hops at the site level. Each local critic has its own parameters. 

\begin{figure}[!htb]
    \centering
    \begin{subfigure}[t]{0.48\linewidth}
        \centering
        \includegraphics[width=\linewidth]{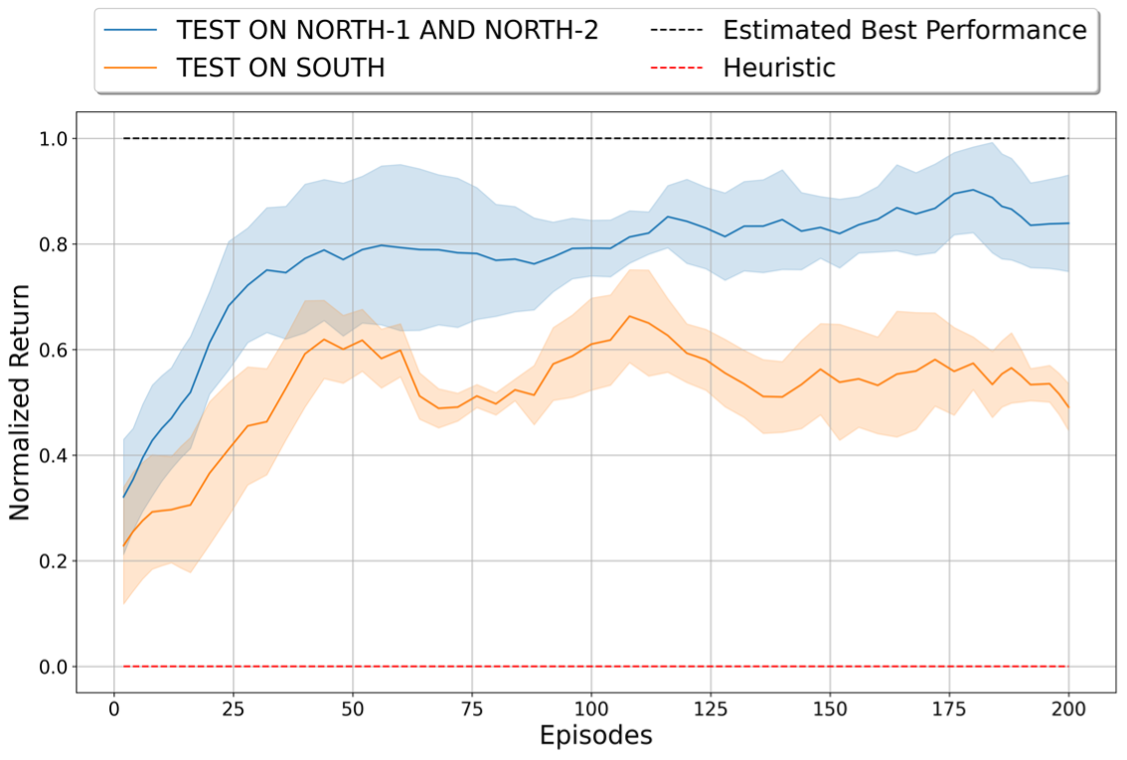}
        \caption{Normalized return on North-$1$, North-$2$, and South regions.}
        \label{fig:gen_return}
    \end{subfigure}\hfill
    \begin{subfigure}[t]{0.48\linewidth}
        \centering
        \includegraphics[width=\linewidth]{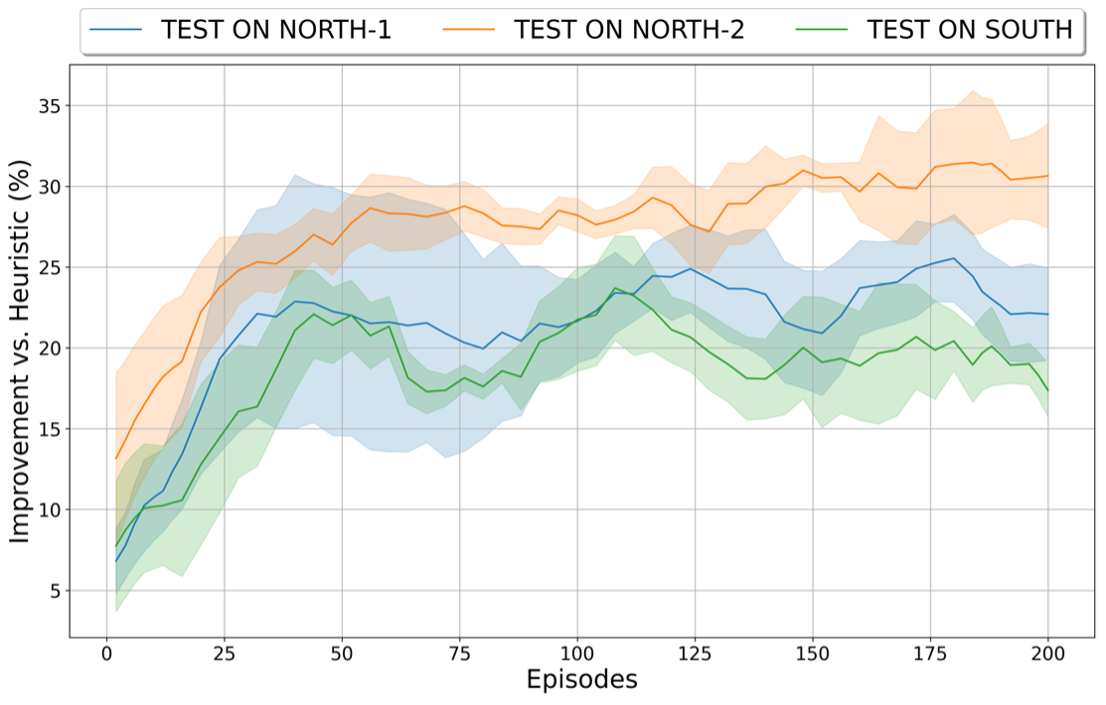}
        \caption{Percentage improvement over heuristic baselines}
        \label{fig:gen_impr}
    \end{subfigure}
    \caption{Generalization results on Manchester.}
    \label{fig:generalization_results}
\end{figure}

\textbf{Generalization under topology shifts.} To evaluate robustness to topology changes, we split the Manchester deployment into North~1, North~2, and South regions, as shown in Fig. \ref{fig:manchester_rem}.
The model is trained on North-$1$ and North-$2$ and evaluated both on the training regions and on South, which is larger and exhibits a different topology. Fig. \ref{fig:generalization_results} reports the generalization performance. The left panel shows normalized expected return to account for differences in scenario size. The right panel reports relative improvement over heuristic baselines. Despite the increased scale and different topology, the learned policy generalizes effectively to the unseen South region.

\textbf{Critics comparison.} Fig. \ref{fig:critics_comparison} compares the three critic designs on the Manchester scenario. The distributed $N$-hops critic consistently achieves faster convergence and higher final performance than both centralized alternatives. Compared to the centralized global-return critic, distributed training improves learning stability without sacrificing performance. On the other hand, when compared to the centralized local-returns critic (site-level outputs), the distributed $M$-hops critic achieves higher network-level throughput, indicating that sub-network level optimization better approximates the global objective than purely site-level optimization.

\section{Conclusion}
\label{sec:conclusion}
This paper addressed the problem of HO in cellular networks through adaptive tuning of CIOs. Motivated by the limitations of rule-based and centralized RL control, we formulated CIO-based HO optimization as a cooperative MARL problem under CTDE. By placing agents on the edges of the cell graph and operating on a dual-graph representation, our formulation aligns the learning architecture with the true locus of control in 3GPP HO procedures. We introduced TD3-D-MA, a dual graph-based MARL algorithm that combines a shared GNN actor with a set of region-wise  critics, enabling decentralized execution using only local KPI information while leveraging global feedback during training. Extensive experiments in a realistic \texttt{ns-3}-based simulator demonstrate that the proposed approach consistently outperforms heuristic and centralized RL baselines. In particular, the dual-graph GNN actor exhibits strong generalization across unseen network topologies, while maintaining learning stability and scalability. 

Interesting directions for future work include $(1)$ extending the framework to jointly optimize multiple HCPs; $(2)$ integrating NR-specific features such as beam-level mobility rewards; $(3)$ exploring distributed critic architectures with federated and hierarchical learning designs, which could enable scalable training across geographically distributed network segments. 

\bibliographystyle{IEEEtran}
\bibliography{biblio_shortened}
\end{document}